\documentclass[twocolumn, aps, superscriptaddress, prb,longbibliography]{revtex4-1}
\usepackage{graphicx,color,dcolumn}
\usepackage{amsmath,amssymb}
\usepackage{bm}
\usepackage{url}

\newcommand{\taux}{\tau_{x}}

\newcommand{\tauz}{\tau_{z}}

\newcommand{\sigmaz}{\sigma_{z}}

\newcommand{\va}{\bm{a}}
\newcommand{\vk}{\bm{k}}

\newcommand{\vB}{\bm{B}}

\newcommand{\vp}{\bm{p}}

\newcommand{\ket}[1]{|{#1}\rangle}
\newcommand{\bra}[1]{\langle{#1}|}

\begin{document}

\preprint{APS/123-QED}

\title {Bulk-edge and bulk-hinge correspondence in inversion-symmetric insulators}
\author {Ryo Takahashi}
\affiliation{
Department of Physics, Tokyo Institute of Technology, 2-12-1 Ookayama, Meguro-ku, Tokyo 152-8551, Japan\\
}
\author {Yutaro Tanaka}
\affiliation{
Department of Physics, Tokyo Institute of Technology, 2-12-1 Ookayama, Meguro-ku, Tokyo 152-8551, Japan\\
}
\author {Shuichi Murakami}
\affiliation{
Department of Physics, Tokyo Institute of Technology, 2-12-1 Ookayama, Meguro-ku, Tokyo 152-8551, Japan\\
}
\affiliation{
TIES, Tokyo Institute of Technology, 2-12-1 Ookayama, Meguro-ku, Tokyo 152-8551, Japan\\
}

\date{\today}

\begin{abstract}
We show that a slab of a three-dimensional inversion-symmetric higher-order topological insulator (HOTI) in class A is a 2D Chern insulator, and that in class AII is a 2D $Z_2$ topological insulator. We prove it by considering a process of cutting the three-dimensional inversion-symmetric HOTI along a plane, and study the spectral flow in the cutting process. 
We show that the $Z_4$ indicators, which characterize three-dimensional inversion-symmetric HOTIs in classes A and AII, are directly related to the $Z_2$ indicators for the corresponding two-dimensional slabs with inversion symmetry, i.e. the Chern number parity and the $Z_2$ topological invariant, for classes A and AII respectively. The existence of the gapless hinge states is understood from the conventional bulk-edge correspondence between the slab system and its edge states. Moreover, we also show that the spectral-flow analysis leads to another proof of the bulk-edge correspondence in one- and two-dimensional inversion-symmetric insulators.  \end{abstract}

\maketitle

\section{introduction}

Bulk-edge correspondence is one of the key concepts in the field of a topological insulator (TI)\cite{RevModPhys.82.3045,RevModPhys.83.1057}, which associates the nontrivial topology of the bulk wave functions with existence of anomalous surface states. 
In the quantum Hall insulator (QHI), which is characterized by the bulk Chern number\cite{PhysRevLett.49.405}, chiral edge modes appear at the edges of the system\cite{PhysRevB.23.5632,PhysRevLett.71.3697}. 
In the quantum spin Hall insulator\cite{PhysRevLett.95.146802,PhysRevLett.95.226801,PhysRevLett.96.106802}, which is the time-reversal-symmetric counterpart of the QHI and characterized by the $Z_2$ topological invariant, helical edge modes appear at the edges of the system. 

In inversion-symmetric insulators, inversion parities at time-reversal-invariant momenta (TRIM) are topological invariants. When they take nontrivial values, the insulator shows unusual properties at the edges. In one-dimensional centrosymmetric  insulators, the Zak phase is quantized, and its value is evaluated from the inversion parities\cite{PhysRevLett.62.2747}. As the Zak phase is proportional to polarization, in the system with the nonzero Zak phase, fractional surface charges appear at the end of the system\cite{PhysRevB.47.1651,PhysRevB.48.4442,Miert_2016}.
In two-dimensional centrosymmetric insulators, the Chern number parity is evaluated from the inversion parities
\cite{PhysRevB.83.245132,PhysRevB.85.165120,PhysRevB.86.115112}. When the Chern number parity is odd, an odd number of chiral edge modes appear at the edges of the system. 

In recent years, a new class of a TI, called higher-order topological insulator (HOTI)
\cite{PhysRevLett.108.126807,
PhysRevLett.110.046404,
PhysRevLett.111.047006,
PhysRevB.89.224503,
PhysRevB.95.165443,
Benalcazar61,
fangc,
PhysRevLett.119.246401,
PhysRevLett.119.246402,
PhysRevB.96.245115,
serra2018observation,
PhysRevLett.120.026801,
peterson2018quantized,
PhysRevB.97.155305,
PhysRevB.97.241402,
PhysRevB.97.205135,
PhysRevB.97.205136,
schindler2018higherTI,
PhysRevB.98.045125,
schindler2018higher,
PhysRevB.98.081110,
PhysRevX.8.031070,
imhof2018topolectrical,
PhysRevB.98.205129,
PhysRevB.98.245102,
PhysRevX.9.011012
}, 
has been proposed. A three-dimensional HOTI is insulating both in the bulk and in the surface. However, it has one-dimensional anomalous gapless states at the hinges, which are intersections of two surfaces. 
The correspondence between the bulk topology and the anomalous gapless states at the hinges is called bulk-hinge correspondence, which is an extension of the conventional bulk-edge correspondence. 

To the best of our knowledge, in the previous studies, explanations of bulk-hinge correspondence are roughly classified into two: 
(i) $\vk\cdot\vp$ theory approach\cite{fangc,PhysRevLett.119.246401,PhysRevLett.119.246402,PhysRevB.97.205135,PhysRevB.97.205136,schindler2018higherTI,schindler2018higher,PhysRevX.8.031070,PhysRevB.98.245102,PhysRevX.9.011012}, 
and (ii) Wannier approach\cite{PhysRevLett.119.246402,PhysRevB.96.245115,PhysRevB.98.081110,PhysRevB.98.245102}. 
In (i), one starts from the surface Dirac Hamiltonian, which represents anomalous gapless surface states as a low-energy effective Hamiltonian for the surface. By adding a symmetry-respecting mass term, the surface energy spectrum becomes gapped. However, on the surface of the HOTI, the sign of the mass term depends on the surface direction. Therefore, at the hinge, shared by two surfaces with opposite signs of the mass terms, the mass term is zero. It means that the hinge states remain gapless. 
In (ii), we consider the two two-dimensional high-symmetry subspaces (e.g. $k_z=0$ and $k_z=\pi$) in the three-dimensional $\vk$-space as pseudo two-dimensional systems, and calculate the Wannier centers (WCs) for the states in these subspaces. 
If excess corner charges at the two subspaces calculated from the WCs are different, there exist hinge states which compensate for the difference of the corner charges.

Both approaches have some drawbacks. Both of them are proofs only for special models, and it is not clear whether they can be applied to general systems. 
First, the $\vk\cdot\vp$ theory approach cannot be applied to systems whose surfaces are not described by the Dirac model. Therefore, one cannot conclude existence of hinge states for general systems from this argument. 
Second, in the WC approach, the exact value of the excess corner charge is concluded only in the case of completely localized Wannier states, which have no hopping terms. 
Therefore, in order to complete the proof, it is necessary to relate the value of the excess corner charge with topological invariants for general systems.

In this paper, we introduce another approach, which is applicable to more general tight-binding models with inversion symmetry.  
We introduce an open boundary to a 3D inversion-symmetric HOTI via cutting procedure\cite{PhysRevB.78.045426}. 
In this process, the boundary condition continuously changes from the periodic one to the open one, while maintaining inversion symmetry. By examining the spectral flow in the cutting process, we obtain restrictions on the inversion parities of the resulting 2D slab system. Thus, in this paper, we show that the $Z_4$ indicator $\mu_1$, which characterizes 3D inversion-symmetric HOTIs in class A, is directly related to the Chern number parity of the resulting 2D slab system in class A. Likewise, we also show that the $Z_4$ indicator $\kappa_1$, which characterizes 3D inversion-symmetric HOTIs in class AII, is directly related to the $Z_2$ topological invariant of the resulting 2D slab system in class AII.
If we cut the 3D HOTI via the cutting procedure, the corresponding 2D topological invariant of the resulting 2D system always takes a nonzero value, as we show in this paper. Due to the conventional bulk-edge correspondence, the resulting 2D system has an odd number of anomalous edge modes at the edges. Since the edges of the resulting 2D system are nothing but the hinges of the original 3D system, this gives a proof of the bulk-hinge correspondence.

Moreover, we apply the same approach to one- and two-dimensional inversion-symmetric topological insulators. As a result, we obtain another proof of bulk-edge correspondence. In the previous papers, inversion parities are related to bulk topological invariants, such as the Zak phase and the Chern number parity. In our approach, inversion parities are directly related to unusual surface properties, such as the fractional end charge and the chiral edge modes. 

We note that, in this paper, we consider a process of cutting the three-dimensional HOTI along a plane. 
In another paper, we consider a process of cutting the three-dimensional HOTI along two planes\cite{Tanaka_preparation}. In the paper, we discuss allowed positions of the hinge states, which is determined only if we consider a cutting along two planes. Thus, different cutting procedure reveal different aspects of hinge states.

This paper is organized as follows. In Sec.~I\hspace{-.1em}I, we introduce the cutting procedure and study the spectral flow in the cutting process. As a result, we obtain another proof of bulk-edge correspondence in a 1D inversion-symmetric topological insulator. In Sec.~I\hspace{-.1em}I\hspace{-.1em}I, by using the result of Sec.~I\hspace{-.1em}I, we show another proof of the famous relation between the number of chiral edge modes and the inversion parities at time-reversal-invariant momenta (TRIM). 
In Sec.~I\hspace{-.1em}V, we show that $\mu_1$, the indicator of 3D inversion-symmetric HOTI, is directly connected to the 2D indicator, i.e. Chern number parity. 
In Sec.~V, we generalize the results of Sec.~\mbox{I\hspace{-.1em}I}-\mbox{I\hspace{-.1em}V} to time-reversal symmetric systems, i.e. class AII systems, and show that $\kappa_1$ is directly connected to the $Z_2$ topological invariant. Conclusion are given in Section V\hspace{-.1em}I.

\section{Bulk-edge correspondence in 1D class A}
We start with a noninteracting centrosymmetric system on a one-dimensional lattice with periodic boundary conditions. 
In such a system, $n_{\pm}(\Gamma_j)$, the number of occupied states with even ($+$) and odd ($-$) parities at a TRIM $\Gamma_j$ are topological invariants. 
Particularly for insulators, their sum is equal to the number of occupied bands $\nu$, i.e. $n_+(\Gamma_j)+n_-(\Gamma_j)=\nu\ (^\forall \Gamma_j)$. 
In one-dimensional systems, there are two TRIM, $k_1=0,\pi$, where $k_1$ is the momentum. Therefore, in an inversion-symmetric insulator, we have three independent topological numbers ($\nu,n_-(0),n_-(\pi)$). If one of these numbers is different between two insulators, they are topologically different in that they cannot be continuously deformed to each other without gap closing or breaking inversion symmetry.

Here, we briefly discuss a choice of the unit cell and the inversion center. We take the unit cell to be invariant under the inversion operation. 
Then, there are two inequivalent inversion centers for the unit cell; one is at the center of the unit cell and the other at the boundary of the unit cell, and they are displaced by a half of the primitive lattice vector.  
In this section, we assume that the number of unit cells, $L$, is odd, and we choose the inversion center to be the center of the unit cell. 
As we explain later in Appendix B, when $L$ is even, we should change the choice of the inversion center. We also note that our unit-cell choice cannot be applied to some models, as we explain at the end of this section.

In this section, we show that for a centrosymmetric one-dimensional system with an open boundary condition, when $n_-(0)-n_-(\pi)=1$ $(\textrm{mod}\ 2)$, the number of occupied states ($N_{\textrm{open}}$) have a different parity from that of bulk ($N_{\textrm{bulk}}$), i.e. $N_{\textrm{open}}=N_{\textrm{bulk}}+1$ $(\textrm{mod}\ 2)$. More precisely, we show the following relation:
\begin{align}
N_{\textrm{open}}=N_{\textrm{bulk}}+n_-(0)-n_-(\pi)\quad(\textrm{mod}\ 2).  \label{nuopen_nubulk}
\end{align}
As we will mention later, this result is closely related to a fractional end charge. 

\subsection{General proof}
\begin{figure}[t]
  \centerline{\includegraphics[width=8cm,clip]{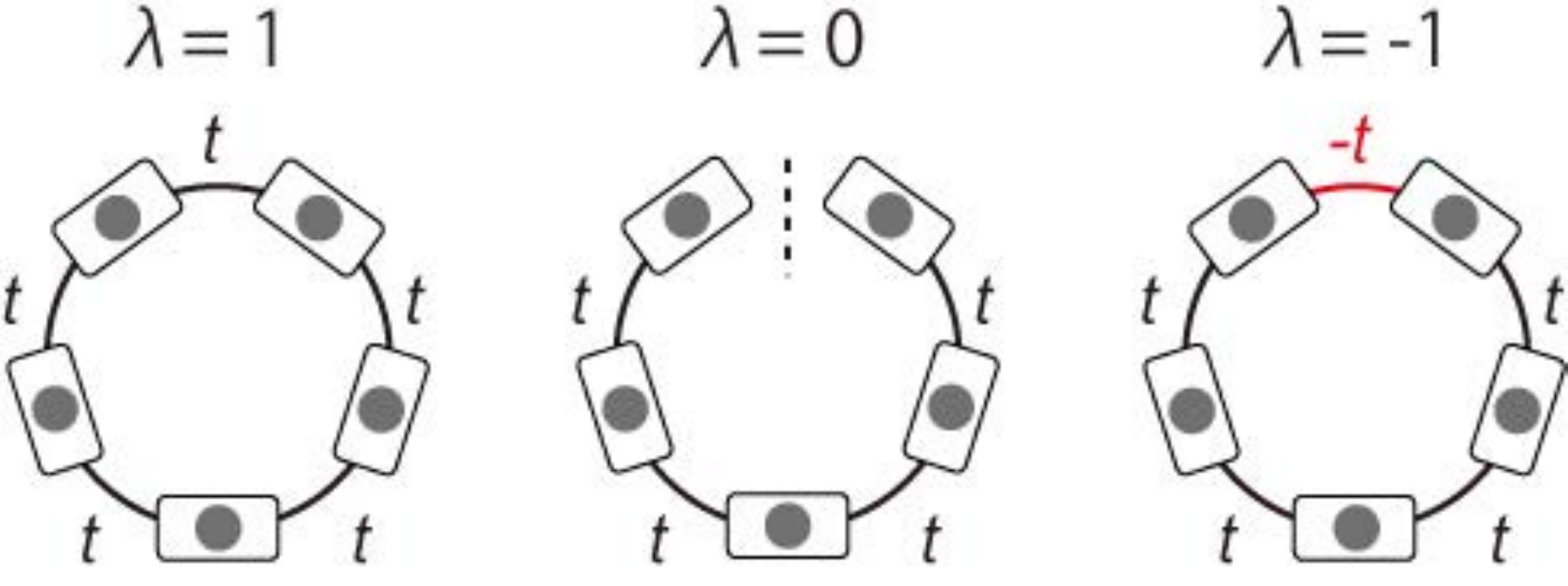}}
  \caption{(Color online) Conceptual figure of a one-dimensional inversion-symmetric insulator cut at the boundary. When $\lambda=1$, the system is periodic, and when $\lambda=0$ the system is open.  When $\lambda=-1$, the system is anti-periodic. 
}
    \label{parity_exchange_1d_A}
\end{figure}

\begin{figure}[t]
  \centerline{\includegraphics[width=8cm,clip]{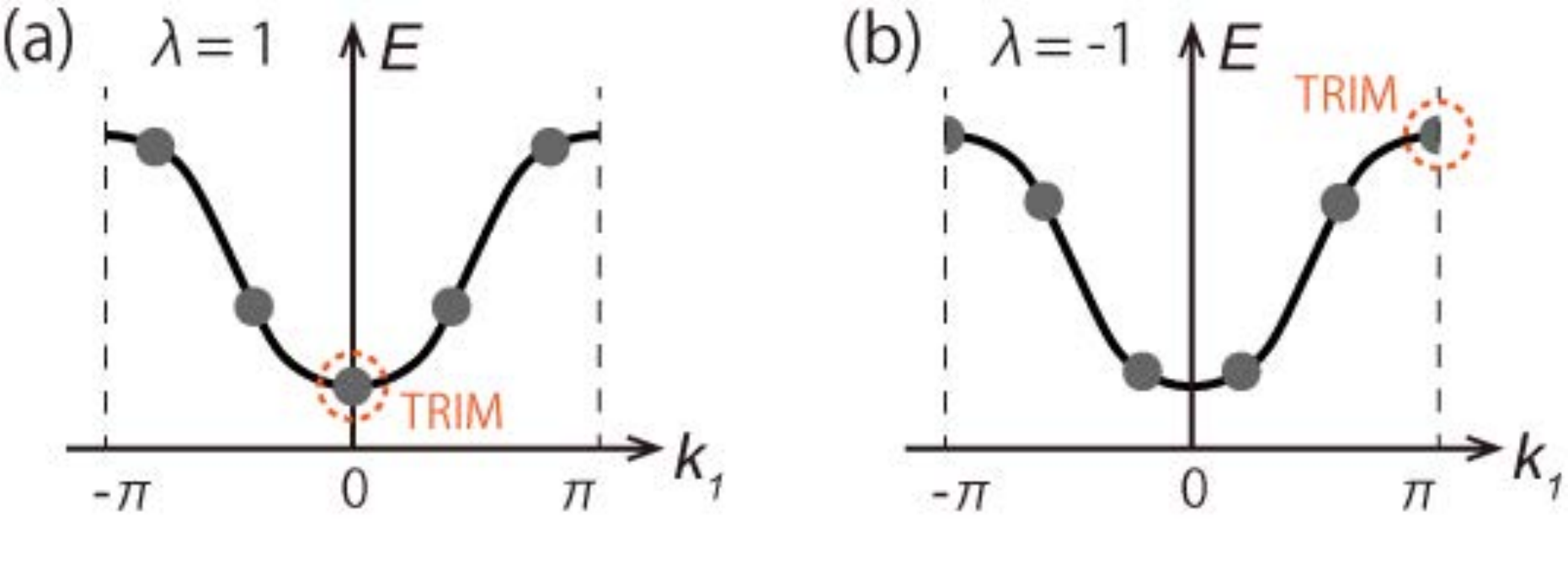}}
  \caption{(Color online) Two examples of the bulk energy levels at $\lambda=\pm1$. All of the states except for those at TRIM are paired by inversion operation. (a) For $\lambda=+1$, the state at $k_1=0$ is not paired. (b) For $\lambda=-1$, the state at $k_1=\pi$ is not paired.
}
    \label{parity_exchange_1d_B}
\end{figure}

Here, we assume that no atomic site is located at the unit-cell boundary. 
We set the system size to be $L|\va|$, where $\va$ is the primitive lattice vector. For simplicity, let $L$ be an odd integer, that is, $L=2M+1$ with an integer $M$. 
The case of even $L$ is discussed in Appendix B. For each unit cell, we associate a lattice site with the middle of the unit cell. 
Let $x_1$ be the coordinate along the 1D system, with $x_1=-M, -M+1, \cdots, M-1, M$ being the lattice sites for the tight-binding model. 
We first begin with a periodic chain, by connecting between the unit cells at the two sites $x=\pm M$. 
Next, in the following way, we introduce a cutting procedure, which is used in appendix of Ref.~[\onlinecite{PhysRevB.78.045426}] for class AII systems. 
We replace the hopping amplitudes $t_j$ for all the bonds that cross the boundary between $x_1=-M$ and $x_1=M$ by $\lambda t_j$, where $\lambda$ is real. 
We note that if $\lambda=1$ the system is periodic in $x_1$, and if $\lambda=0$ the system is open in the $x_1$ direction. 
Figure \ref{parity_exchange_1d_A} is a conceptual figure of the one-dimensional inversion-symmetric insulator cut at the boundary. 
For any real values of $\lambda$, inversion symmetry is always preserved. 
Let $N_{-(+)}$ be the number of occupied states with odd (even) parity; these values depend on $\lambda$. 
First of all, we show that the value of $N_-$ is evaluated from $n_{-}(\Gamma_j)$ for some specific values of $\lambda$. 

First, we consider the case of $\lambda=1$, when the system is periodic in $x_1$. In this case, the Bloch wave number $k_1$ takes the following values: 
\begin{align}
k_1=\frac{2\pi}{L}m_1\quad (-M\leq m_1\leq M,\ M:\text{integer}).
\end{align}
Because $L\ (=2M+1)$ is an odd number, $k_1$ is a TRIM if and only if $k_1=0$, as shown in Fig.~\ref{parity_exchange_1d_B}(a). 
For other values of $k_1=k_{*}(\neq0)$, the inversion operation changes $k_{*}$ to $-k_{*}$. One can construct eigenstates of the inversion operator $\mathcal{P}$ with eigenvalues $+1$ and $-1$ from $\psi_m(k_*)$ and $\psi_m(-k_*)$ as $\frac{1}{\sqrt{2}}(\psi_m(k_*)\pm \mathcal{P}\psi_m(k_*))$, where $\psi_m(k)$ is a Bloch eigenstate and $\mathcal{P}\psi_m(k_*)\propto\psi_m(-k_*)$. 
Therefore, each pair $(k_*,-k_*)$ contributes 1 to $N_-$. 
For an insulating system, the total number of non-TRIM pairs is evaluated as $(L-1)\nu/2$, where $\nu$ is the number of occupied bands. 
Therefore, when $\lambda=1$, $N_-$ can be expressed as follows: 
\begin{align}
N_{-}|_{\lambda=1}=\frac{(L-1)\nu}{2}+n_{-}(0). \label{Nn0}
\end{align}

Next, we consider the case of  $\lambda=-1$, when the system is anti-periodic in $x_1$. In this case, by performing a unitary transformation $U_1=\exp[i\pi\hat{x_1}/L]$, where $\hat{x_1}$ is the position operator, the periodicity in the $x_1$ direction can be restored, but the Bloch wave vector is shifted as $k_1\to k_1+\pi/L$ (see Appendix A). Then $k_1$ takes the following values:
\begin{align}
k_1=\frac{2\pi}{L}m_1+\frac{\pi}{L},\quad (-M\leq m_1\leq M).
\end{align}
Now,  as shown in Fig.~\ref{parity_exchange_1d_B}(b), $k_1$ is a TRIM if and only if $k_1=\pi$. Therefore, 
\begin{align}
N_{-}|_{\lambda=-1}=\frac{(L-1)\nu}{2}+n_{-}(\pi). \label{Nnpi}
\end{align}

Next, we follow the change in $N_-$ through the change of $\lambda$ from $1$ to $-1$. 
From Eqs.~(\ref{Nn0}) and (\ref{Nnpi}), the total change in $N_{-}$ is evaluated from $n_-$:
\begin{align}
N_{-}|_{\lambda=1}-N_{-}|_{\lambda=-1}&=n_{-}(0)-n_{-}(\pi). \label{NNnn}
\end{align}
We consider the energy spectrum in the process of changing $\lambda$. In this process, the Hamiltonian changes only at the boundary. 
Therefore, only the boundary localized states have strong $\lambda$ dependence. 
We can show that, for sufficiently large $L$, the energy spectrum is symmetric with respect to the transformation $\lambda\leftrightarrow-\lambda$, and the bound states $\ket{\psi_l(\lambda)}$ and $\ket{\psi_l(-\lambda)}$ have opposite parities (see Appendix B). 
This means that the following relation holds (double sign in the same order): 
\begin{align}
[N_{\pm}]^{\lambda=1}_{\lambda=0}=[N_{\mp}]^{\lambda=-1}_{\lambda=0}. 
\label{NpmNmp}
\end{align}
Here, we used the following notation: $[X]^{\lambda=b}_{\lambda=a}\overset{\mathrm{def}}{=}X|_{\lambda=b}-X|_{\lambda=a}$. 
By using Eqs.~(\ref{NNnn}) and (\ref{NpmNmp}), the difference between $N_{\text{bulk}}\equiv N|_{\lambda=1}$ and $N_{\text{open}}\equiv N|_{\lambda=0}$ is evaluated as follows: 
\begin{align}
[N]^{\lambda=1}_{\lambda=0}
&=[N_+]^{\lambda=1}_{\lambda=0}+[N_-]^{\lambda=1}_{\lambda=0}
\notag \\
&=[N_-]^{\lambda=-1}_{\lambda=0}+[N_-]^{\lambda=1}_{\lambda=0}
\notag \\
&=[N_-]^{\lambda=-1}_{\lambda=1}+2[N_-]^{\lambda=1}_{\lambda=0}.
\end{align}
Therefore, we get 
\begin{align}
N|_{\lambda=0}=N|_{\lambda=1}+n_{-}(0)-n_{-}(\pi)+2[N_-]^{\lambda=0}_{\lambda=1}.
\label{nuopen_nubulk2}
\end{align}
Equation (\ref{nuopen_nubulk}) is derived by taking modulo $2$ on both sides of the Eq.~(\ref{nuopen_nubulk2}).

Here, we explain the relation of this result and the fractional end charge. From Eq.~(\ref{nuopen_nubulk}), when $n_-(0)-n_-(\pi)\equiv1$ (mod 2), through the change from $\lambda=1$ to $\lambda=0$, the number of occupied states changes by an odd number. Since the Hamiltonian is changed only at the boundary, newly occupied states are localized at the boundary, and it causes total excess charges $(2m+1)e$ ($m$: integer) at the two ends of the system. If we assume that the system preserves inversion symmetry, the excess charges should be divided equally to the two ends. Therefore, each end have a fractional excess charge $(m+\frac{1}{2})e$.

\subsection{Illustrative examples}
\begin{figure}[t]
  \centerline{\includegraphics[width=8cm,clip]{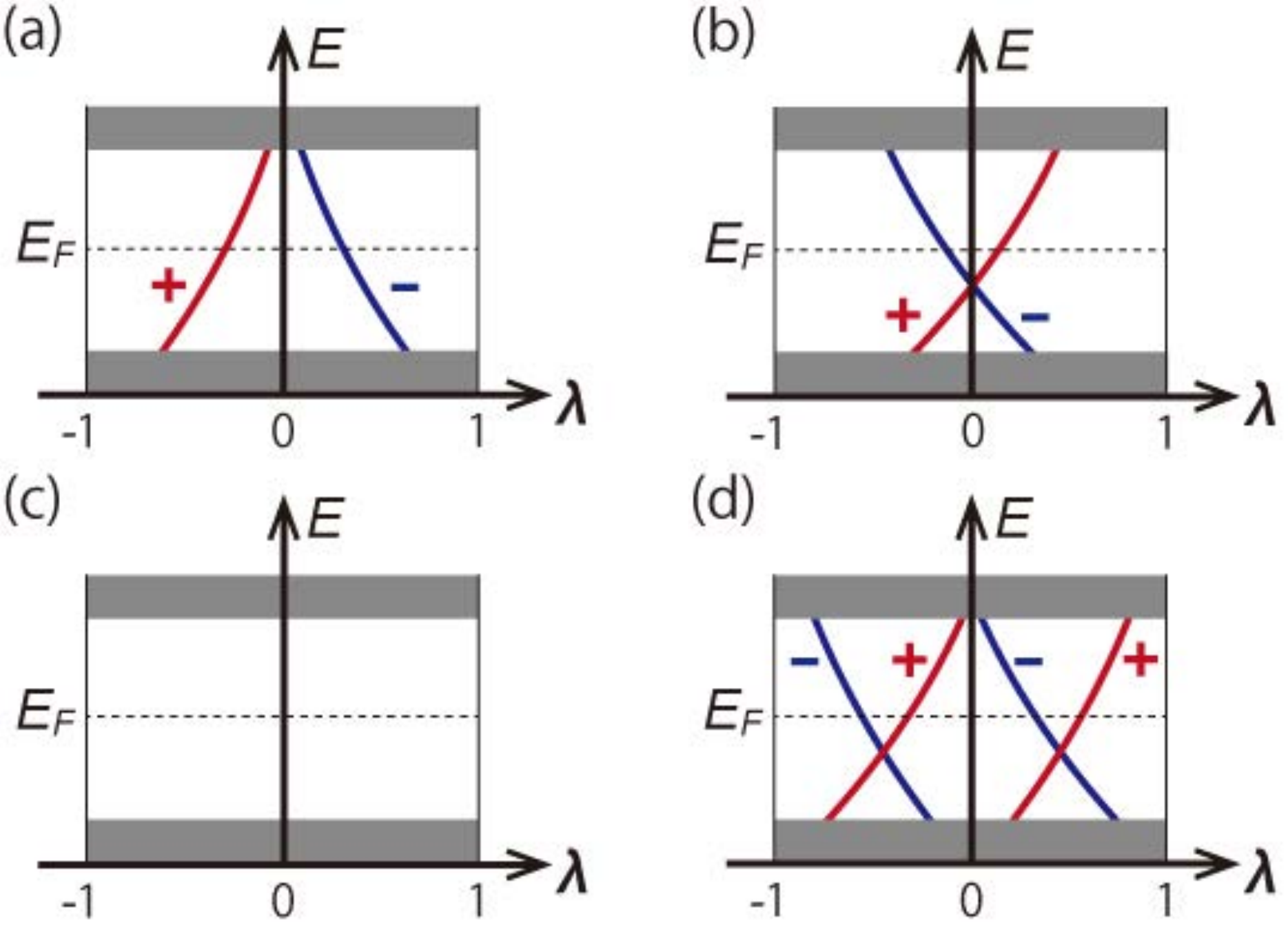}}
  \caption{(Color online) Four representative examples of 
the spectral flow for the change of $\lambda$.Red and blue lines, which are marked with $+$ and $-$, represent bound states with even and odd parity, respectively. The shaded areas represent the bulk energy spectra. 
The number of occupied states with odd parity at TRIM, $(n_{-}(0),n_-(\pi))$, is (a-b) (1,0) (c) $(0,0)$ and (d) $(2,0)$.
In (a-b), $N_{\textrm{open}}$, the number of states below the Fermi energy $E_F$ at $\lambda=0$, is one less or more than that of the bulk, $N_{\textrm{bulk}}$. 
In (c-d), $N_{\textrm{open}}$ and $N_{\textrm{bulk}}$ are equal. 
}
    \label{parity_exchange_1d_2}
\end{figure}

Here, we illustrate how the spectrum changes through the change of $\lambda$ for three examples: 
(i) $(n_{-}(0),n_-(\pi))=(1,0)$, (ii) $(0,0)$ and (iii) $(2,0)$. By considering the spectral flow, we can visualize how Eq.~(\ref{nuopen_nubulk2}) is satisfied. In the case (i), when $\lambda$ is changed from $+1$ to $-1$, one state with odd parity crosses the Fermi energy $E_F$ from below, and one state with even parity crosses $E_F$ from above. Furthermore, the energy spectrum is symmetric with respect to the transformation $\lambda\leftrightarrow-\lambda$, and the bound states $\ket{\psi_l(\lambda)}$ and $\ket{\psi_l(-\lambda)}$ have opposite parities (see Appendix B). 
Figures \ref{parity_exchange_1d_2}(a,b) show two representative examples of the energy spectra which satisfy these requirements. 
In Fig.~\ref{parity_exchange_1d_2}(a), the number of occupied states is one less than that of the bulk, i.e. $N_{\textrm{open}}=N_{\textrm{bulk}}-1$. 
In Fig.~\ref{parity_exchange_1d_2}(b), the number of occupied states is one more than that of the bulk, i.e. $N_{\textrm{open}}=N_{\textrm{bulk}}+1$. 
In both cases $N_{\textrm{open}}$ have different parity from that of $N_{\textrm{bulk}}$. 
In the case (ii) $(n_{-}(0),n_-(\pi))=(0,0)$, the total change in $N_-$ is equal to $0$. Figure \ref{parity_exchange_1d_2}(c) shows an example of the energy spectrum. In Fig.~\ref{parity_exchange_1d_2}(c), there is no state which crosses $E_F$, and $N_{\textrm{open}}=N_{\textrm{bulk}}$.  
Finally, we consider case (iii) $(n_{-}(0),n_-(\pi))=(2,0)$. 
In this case, when $\lambda$ is changed from $+1$ to $-1$, two odd- (even-) parity states cross $E_F$ from below (above). 
One of the examples is shown in Fig.~\ref{parity_exchange_1d_2}(d), and $N_{\textrm{open}}$ is equal to that of the bulk.   
In both cases (ii) and (iii), $N_{\textrm{open}}$ have the same parity with that of $N_{\textrm{bulk}}$. 

\subsection{Assumptions on the unit-cell choice}

\begin{figure}[t]
  \centerline{\includegraphics[width=8cm,clip]{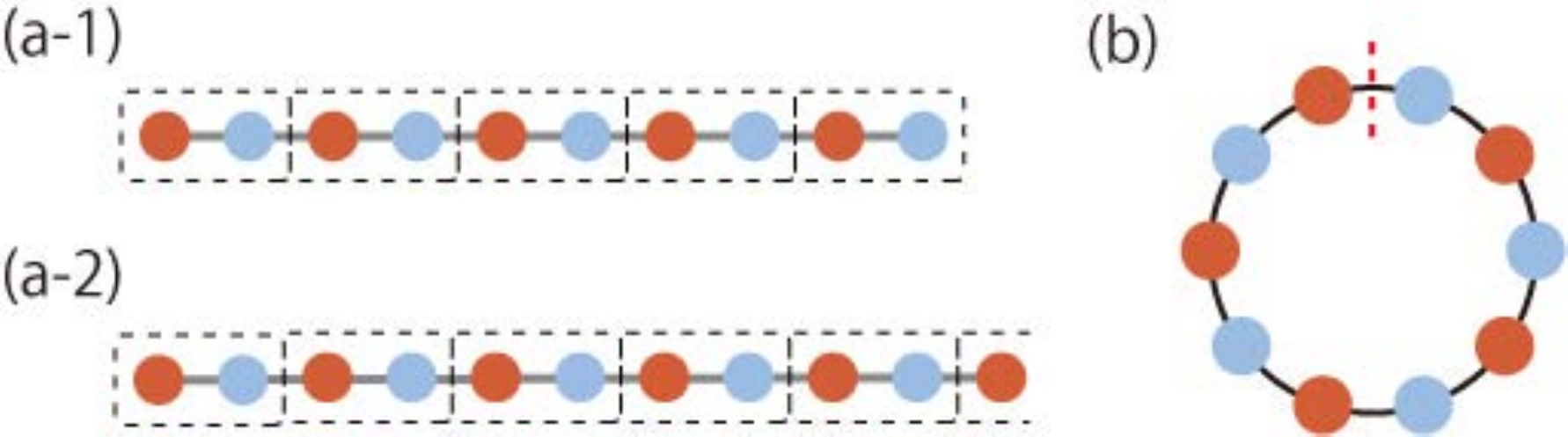}}
  \caption{(Color online) Examples of the chains which cannot be treated within our scenario with (a-1,2) open boundary conditions, and (b) periodic boundary conditions. 
Colored balls represent the atomic sites. The red balls and the blue balls are located at distinct inversion centers. (a-1) is obtained by cutting the periodic chain in (b), not at an inversion center. (a-2) cannot be constructed from the periodic chain in (b), because the number of red balls and that of blue balls are different in (a-2). 
}\label{commensurate_A}
\end{figure}

In our theory, we begin with a periodic chain consisting of an integer number of unit cells, and cut the chain at an inversion center. From this construction, we see that not every open chain can be treated within our theory. For example, the open chain in Fig.~\ref{commensurate_A}(a-1), where blue and red balls represent different atoms, is outside of our theory because the chain is cut not at an inversion center. The open chain in Fig.~\ref{commensurate_A}(a-2) is also outside of our theory because it cannot be constructed from a periodic chain. Therefore, the periodic chain Fig.~\ref{commensurate_A}(b) cannot be treated within our scenario.

Here, we discuss the assumptions on the unit-cell choice used in our proof. 
We have two assumptions: (i) the unit cell is invariant under the inversion operation, and (ii) no atomic site is located at the unit-cell boundary. 
In the following, we briefly explain that this assumption is physically reasonable as long as we only use the knowledge of the bulk inversion eigenvalues. 
We also show an example which cannot meet these two assumptions simultaneously, and it is outside the scope of our proof. 
However, in such an example, the bulk-edge correspondence cannot be understood only from the knowledge of the bulk inversion eigenvalues\cite{PhysRevB.89.161117,PhysRevB.95.035421,PhysRevB.96.235130,PhysRevX.8.021065}. In this sense, our proof fully covers the bulk-edge correspondence detected only from the knowledge of the bulk inversion eigenvalues.

\begin{figure}[t]
  \centerline{\includegraphics[width=8cm,clip]{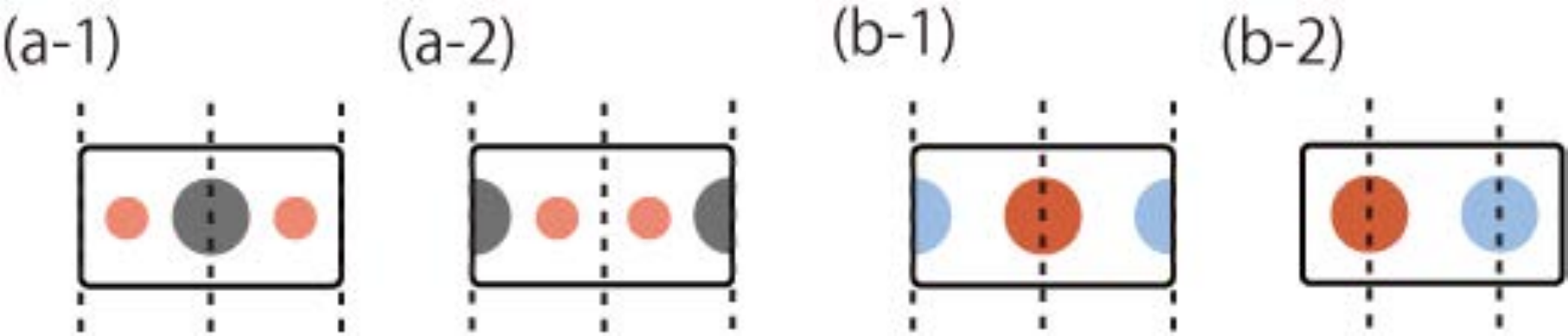}}
  \caption{(Color online) Examples of unit cells. The circles represent atoms. The unit cells of (a-1) and (c-2), (b-1) and (b-2) corresponds to the same periodic system, respectively. The unit cells satisfy the assumption (i) except for (b-2). Only the unit cells (a-1) and (b-2) satisfy the assumption (ii). 
}\label{commensurate_B}
\end{figure}

First, we explain the assumption (ii). 
Figures \ref{commensurate_B}(a-1) and (a-2) show two examples of the unit cell of the same periodic system, whose difference is the choice of the center of the unit cell. Here, the balls represent the atomic sites, and the dashed lines represent the inversion centers.  
In Fig.~\ref{commensurate_B}(a-1), there are no atomic sites at the unit-cell boundary, and the assumption (ii) is satisfied. 
On the other hand, in Fig.~\ref{commensurate_B}(a-2), the atomic site is located at the unit-cell boundary. Since we cannot cut the chain at the atomic site, the unit-cell boundary cannot coincide with the edge of the open system, and this choice of the unit cell in Fig.~\ref{commensurate_B}(a-2) cannot be adopted.

We mentioned that the periodic chain Fig.~\ref{commensurate_A}(b) cannot be treated within our scenario. From the viewpoint of the above two assumptions (i), (ii), we can equivalently say that we cannot choose a unit cell that simultaneously satisfies the two assumptions (i), (ii), as we can see from Figs.~\ref{commensurate_B}(b-1) and (b-2).

Finally, we briefly explain that our assumptions are physically natural for the purpose of obtaining information on the edge state from inversion eigenvalues, in the light of recent works on the Zak phase\cite{PhysRevB.89.161117,PhysRevB.95.035421,PhysRevB.96.235130,PhysRevX.8.021065}. 
In Ref. \onlinecite{PhysRevB.95.035421}, the Zak phase $\gamma_{\text{Zak}}$ is splitted to the intracellular part $\gamma^{\text{intra}}_{\text{Zak}}$ and the intercellular part $\gamma^{\text{inter}}_{\text{Zak}}$, and the latter is proportional to the excess edge charge,
\begin{align}
\gamma_{\text{Zak}}&=\gamma^{\text{intra}}_{\text{Zak}}+\gamma^{\text{inter}}_{\text{Zak}}, 
\\
Q_{\text{acc}}^{L(R)}&=+(-)\frac{e}{2\pi}\gamma^{\text{inter}}_{\text{Zak}}\ (\text{mod}\ e), \label{Q_acc}
\end{align}
where $Q_{\text{acc}}^{L(R)}$ represents the excess edge charge accumulated at the left (right) edge of the insulating open system. 
Equation (\ref{Q_acc}) holds when the assumption (ii) is satisfied. 
The intracellular part, $\gamma^{\text{intra}}_{\text{Zak}}$, corresponds to the electronic part of the classical polarization of the bulk's unit cell. 
Though, the Zak phase $\gamma_{\text{Zak}}$ is independent of the choice of the unit cell, the intra- and intercellular parts of the Zak phase, $\gamma^{\text{intra}}_{\text{Zak}}$ and $\gamma^{\text{inter}}_{\text{Zak}}$, depend on the choice of the unit cell. If the unit cell is taken to be invariant under the inversion symmetry (assumption (i)), $\gamma^{\text{intra}}_{\text{Zak}}$ vanishes, and $\gamma^{\text{inter}}_{\text{Zak}}=\gamma_{\text{Zak}}$. Moreover, if we choose the real-space origin at one of the inversion centers, $\gamma_{\text{Zak}}$ is quantized to 0 or $\pi$ and calculated by the bulk inversion eigenvalues. Therefore,  $Q^{L(R)}_{\text{acc}}$ is calculated by the bulk inversion eigenvalues if the assumptions (i) and (ii) are met. 

\section{bulk-edge correspondence in 2D  class A}
In this section, we show bulk-edge correspondence in two-dimensional inversion-symmetric insulators. 
We consider a noninteracting centrosymmetric system on a two-dimensional lattice with periodic boundary conditions. In a two-dimensional system, there are four TRIM, $(k_1,k_2)=(0,0), (\pi,0), (0,\pi), (\pi,\pi)$, which we label with $\Gamma$, $X$, $Y$, $M$, respectively. Therefore, in an insulator, we have five independent topological numbers ($\nu,n_-(\Gamma_j)$), where $\Gamma_j$ represent the TRIM.

In this section, we show that for an insulating system with an open boundary condition, when $\sum_{\Gamma_j} n_-(\Gamma_j)=1$ $(\textrm{mod}\ 2)$, there are an odd number of chiral edge modes at the boundary. A similar result has been already shown in previous works \cite{PhysRevB.83.245132,PhysRevB.85.165120,PhysRevB.86.115112} in the form, 
\begin{align}
\prod_{\Gamma_j\in\textrm{TRIM}} (-1)^{n_-(\Gamma_j)}=(-1)^{\textrm{Ch}}, \label{parity_chern}
\end{align}
where Ch is the Chern number of the system. 
Since the Chern number is equal to the number of chiral edge modes \cite{PhysRevB.23.5632,PhysRevLett.71.3697}, this equation also means the existence of chiral edge modes when $\sum_{\Gamma_j} n_-(\Gamma_j)=1$ $ (\textrm{mod}\ 2)$. In this sense, we give another proof of this equation. 
We note that, as is the case with for the one-dimensional systems, we take the inversion-symmetric unit cell which is commensurate with the open system. 

\subsection{General proof}
\begin{figure}[t]
  \centerline{\includegraphics[width=8cm,clip]{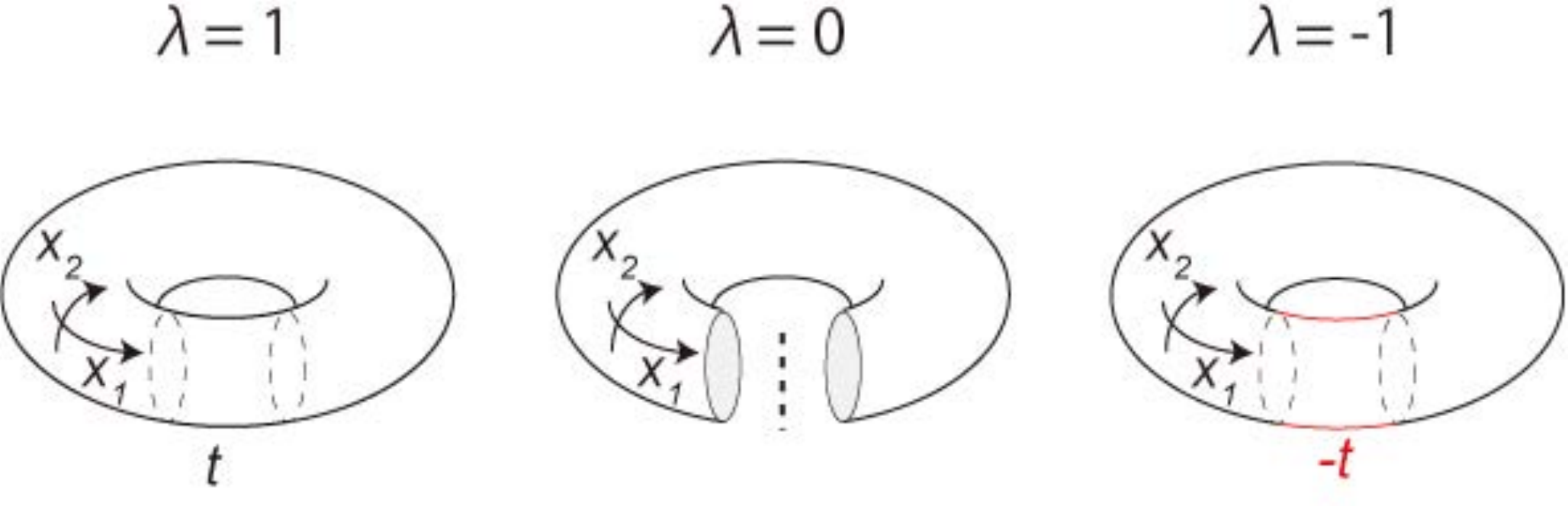}}
  \caption{(Color online) Conceptual figure of a two-dimensional inversion-symmetric insulator cut at the boundary along the $x_1$ direction. When $\lambda=1$, the system is periodic, and when $\lambda=0$ the system is an open finite system at $x_1=$ const.  When $\lambda=-1$, this system is anti-periodic along the $x_1$ direction. 
}\label{2D_boundary_fig}
\end{figure}

We consider a two-dimensional crystal with primitive vectors $\va_i\ (i=1,2)$. Let $L_i\ (i=1,2)$ be the length of the system along $\va_i$, measured in units of $|\va_i|\ (i=1,2)$. For simplicity, let $L_1$ be an odd integer, $L_1=2M_1+1\ (M_1:\text{integer})$, and let $L_2\to\infty$. The centers of the unit cells are located at $x_1=-M_1,-M_1+1,\cdots,M$, measured in the unit of $|\va_1|$. As in the one-dimensional case, we replace the hopping amplitudes $t_j$ for all bonds that cross the boundary between $x_1=-M_1$ and $x_1=M_1$ by $\lambda t_j$, where $\lambda$ is real. Note that we impose a periodic boundary condition along the $x_2$ direction in the rest of this section. Figure \ref{2D_boundary_fig} is a conceptual figure of a two-dimensional inversion-symmetric insulator cut at the boundary between $x_1=-M_1$ and $x_1=M_1$. For real $\lambda$, inversion symmetry is always preserved.

The inversion operator $\mathcal{P}$ changes the 2D wave vector $\vk=(k_1,k_2)$ to $(-k_1,-k_2)$. 
Let us focus on the 1D $\mathcal{P}$-invariant subspaces $k_2=0$ and $k_2=\pi$, which we call $\overline{\Gamma}$ and $\overline{Y}$, respectively. We can regard $\overline{\Gamma}$ and $\overline{Y}$ as 1D TRIM. 
Both of the subspaces can be considered as an effective 1D inversion-symmetric system. 
Let $N_{\overline{\Gamma}}$ and $N_{\overline{Y}}$ be the numbers of occupied states at $k_2=0$ and that of $k_2=\pi$, respectively. 
From Eq.~(\ref{nuopen_nubulk}), the change of $N_{\overline{\Gamma}}$ and $N_{\overline{Y}}$ (mod 2) by changing $\lambda=1$ to $\lambda=0$ is evaluated from $n_-(\Gamma_j)$ ($\Gamma_j\in\textrm{2D TRIM}$): 
\begin{align}
N_{\overline{\Gamma}}|_{\lambda=0}\equiv N_{\overline{\Gamma}}|_{\lambda=1}+n_-(\Gamma)-n_-(X) \quad (\textrm{mod 2}), 
\label{nu0gx}
\\
N_{\overline{Y}}|_{\lambda=0}\equiv N_{\overline{Y}}|_{\lambda=1}+n_-(Y)-n_-(M) \quad (\textrm{mod 2}). 
\label{nupiym}
\end{align}
Since the system is insulating for $\lambda=1$, we get $N_{\overline{\Gamma}}|_{\lambda=1}=N_{\overline{Y}}|_{\lambda=1}=L_1\nu$. 
By subtracting Eq.~(\ref{nupiym}) from Eq.~(\ref{nu0gx}), we obtain the following equation: 
\begin{align}
N_{\overline{\Gamma}}|_{\lambda=0}-N_{\overline{Y}}|_{\lambda=0}
&\equiv  n_-(\Gamma)-n_-(X)-n_-(Y)+n_-(M) \notag \\
&\equiv \sum_{\Gamma_j\in\textrm{TRIM}} n_-(\Gamma_j) \quad (\textrm{mod 2}). \label{nu_chern}
\end{align}
From Eq.~(\ref{nu_chern}), when $\sum n_-(\Gamma_j)=1$ (mod 2), for the system with an open boundary condition, the parity of $N_{\overline{\Gamma}}$ is different from that of $N_{\overline{Y}}$. 
In order to compensate the difference, 
an odd number of chiral edge modes must exist between $\overline{\Gamma}$ and $\overline{Y}$. 

Moreover as pointed out in [\onlinecite{PhysRevB.78.045426}], inversion parities at the respective TRIM have more information than their sum over all the TRIM. For example, if $\sum_{\Gamma_j}n_-(\Gamma_j)=1$, we have two possibilities 
\begin{align}
([N_{\overline{\Gamma}}]^{\lambda=0}_{\lambda=1},
[N_{\overline{Y}}]^{\lambda=0}_{\lambda=1})
\equiv
(1,0), (0,1)
\quad (\textrm{mod 2}),
\end{align}
and we can identify which possibility is realized from parity eigenvalues $n_-(\Gamma_j)$, from Eqs.~(\ref{nu0gx}) and (\ref{nupiym}).  
This indicates which of the 1D TRIM, $\overline{\Gamma}$ and $\overline{Y}$, is ``inside of the boundary Fermi surface''. Here, we define ``inside of the boundary Fermi surface'' as a subspace of $k$-space where the number of occupied states is more or less than that of the bulk by an odd number. 
In order to express this feature, we define boundary fermion parity at $k_y$ to be $n_{\text{boundary}}(k_y)=[N_{k_y}]^{\lambda=0}_{\lambda=1}$, as an analogue of the surface fermion parity in [\onlinecite{PhysRevB.78.045426}]. 
This boundary fermion parity $n_{\text{boundary}}(k_y)$ is always an integer. 
From (\ref{nu0gx}) and (\ref{nupiym}), we define the boundary fermion parity at $\overline{\Gamma_j}$ to be $\sum_{\Gamma_i}n_-(\Gamma_i)$ (mod 2) where the sum is taken over the 2D TRIM $\Gamma_i$ which is projected onto $\overline{\Gamma_j}$ (= $\overline{\Gamma},\overline{Y}$). 
As long as inversion symmetry is preserved, when we modify the system perturbatively without closing the gap, $n_{\text{boundary}}(\overline{\Gamma_j})$ changes by an even number. Therefore, the parity of $n_{\text{boundary}}(\overline{\Gamma_j})$ is a topological number, and it is natural that it is given by the parities at the 2D TRIM.

Here, we note the relation of our study and the previous works. Previous works\cite{PhysRevB.83.245132,PhysRevB.85.165120,PhysRevB.86.115112} have shown that, for two-dimensional systems, the quantity $(-1)^{\text{Ch}}$, where Ch is the Chern number along a crystal plane of the system, is equal to the product of the inversion parities at TRIM on the plane in $\vk$-space. 
Other studies\cite{PhysRevB.23.5632,PhysRevLett.71.3697} have shown that, for two-dimensional systems, the number of chiral edge modes is equal to the Chern number of the system. By combining these two results, we conclude that the quantity $(-1)^{N_{\text{edge}}}$, where $N_{\text{edge}}$ is the number of chiral edge modes, is equal to the product of the inversion parities at TRIM. This is exactly the same statement which we have shown in this section. 
However, compared to the previous studies on bulk-edge correspondence \cite{PhysRevB.23.5632,PhysRevLett.71.3697}, our proof directly relates the bulk inversion parities to the existence of the chiral edge modes. 
Moreover, our proof clarifies that the inversion parities at each TRIM have more information than their product \cite{PhysRevB.83.245132,PhysRevB.85.165120,PhysRevB.86.115112}, leading to the notion of boundary fermion parity $n_{\text{boundary}}(\overline{\Gamma_j})$. 
Therefore, our proof gives a new perspective to the existing results.

\subsection{Illustrative examples}
\begin{figure}[t]
  \centerline{\includegraphics[width=8cm,clip]{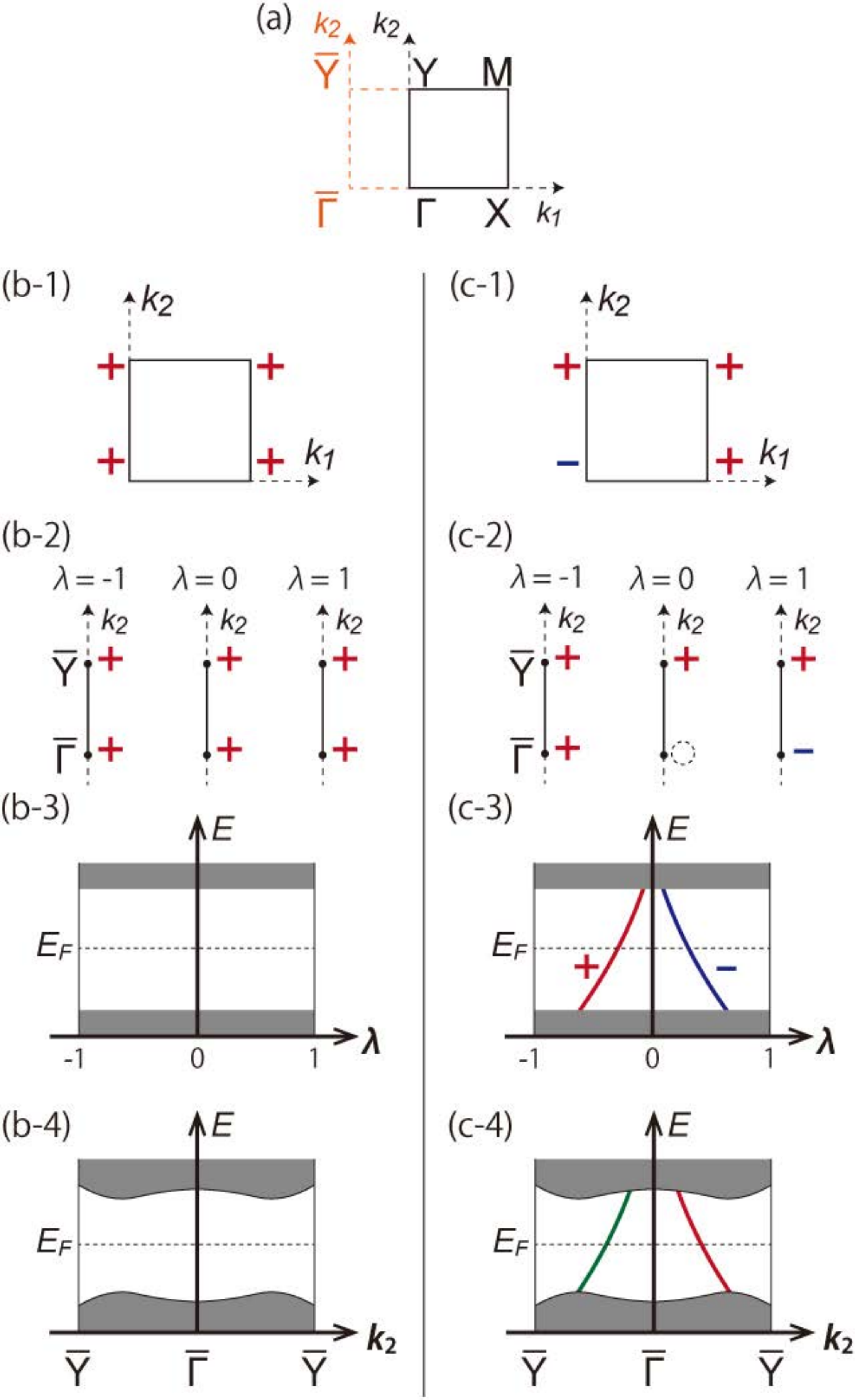}}
  \caption{(Color online) Two representative examples of bulk parities and corresponding energy spectra. (a) Quarter of the 2D Brillouin zone with the symmetry labels shown in black, and a half of the projected 1D Brillouin zone with the symmetry labels shown in orange. (b) (c): Two examples of inversion parities and resulting spectra. (b-1)-(b-4) and (c-1)-(c-4) show two different cases. 
(b-1,c-1) Inversion parities at 2D TRIM. 
(b-2,c-2) Parity arrangements at $\overline{\Gamma}$ and $\overline{Y}$ for $\lambda=-1,0,1$. 
In addition to the state whose parity is shown in (b-2) and (c-2), there are $\frac{(L_1-1)\nu}{2}$ states with even parity and $\frac{(L_1-1)\nu}{2}$ states with odd parity, and they are not shown in the figure for simplicity. In (c-2), at $\lambda=0$, the parity at $\overline{\Gamma}$ is blank, because the number of occupied states is one less than that of the bulk.  
(b-3,c-3) Energy spectra at $\overline{\Gamma}$ in the process of changing $\lambda$. 
(b-4,c-4) Energy spectrum. In (c-4), $N_{\overline{\Gamma}}\neq N_{\overline{Y}}$ (mod 2), and an odd number of chiral edge modes exist. 
}
    \label{parity_exchange_2d}
\end{figure}

Here, we show two examples to see how the system evolves by the change of $\lambda$. Figure \ref{parity_exchange_2d} shows two examples of the bulk parity and corresponding surface energy spectrum. Figure \ref{parity_exchange_2d}(b-1) shows the case with $\sum n_-(\Gamma_j)=0$. In this case, inversion parities at $\overline{\Gamma}$ and $\overline{Y}$ do not depend on $\lambda$, as shown in Fig.~\ref{parity_exchange_2d}(b-2) and (b-3). Therefore, the corresponding surface energy spectrum, Fig.~\ref{parity_exchange_2d}(b-4), does not have a chiral edge mode. 
Figure \ref{parity_exchange_2d}(c-1) shows the case with $\sum n_-(\Gamma_j)=1$. In this case, inversion parities at $\overline{\Gamma}$ depend on $\lambda$, as shown in Fig.~\ref{parity_exchange_2d}(c-2) and (c-3). Therefore, when $\lambda=0$, the number of states at $\overline{\Gamma}$ is one less than that of bulk. On the other hand, inversion parities at $\overline{Y}$ do not depend on $\lambda$ as shown in Fig.~\ref{parity_exchange_2d}(c-2). 
Therefore, the corresponding surface energy spectrum, Fig.~\ref{parity_exchange_2d}(c-4), have a chiral edge mode, which compensates the difference of the number of occupied states at $\overline{\Gamma}$ and $\overline{Y}$.

\section{bulk-hinge correspondence in 3D  class A}
In this section, we consider a three-dimensional noninteracting centrosymmetric insulator with periodic boundary conditions. In a three-dimensional system, there are eight TRIM. As in the case of one- and two-dimensional insulators, we have nine independent topological numbers ($\nu,n_-(\Gamma_j)$), where $\nu$ is filling, and $n_-(\Gamma_j)$ is the number of occupied states with odd parity at the TRIM $\Gamma_j$. 
Interestingly, some combinations of $\{n_-(\Gamma_j)\}$ cannot be realized in atomic insulators. 
The following four numbers, calculated from $n_-(\Gamma_j)$, are indicators which specify whether the combinations of $\{n_-(\Gamma_j)\}$ can be realized in an atomic insulator or not\cite{po2017symmetry,bradlyn2017topological_nature,PhysRevB.98.115150}:
\begin{align}
\mu_1&=\sum_{\Gamma_j:\textrm{TRIM}}\frac{n_+(\Gamma_j)-n_-(\Gamma_j)}{2}
\notag \\
&\equiv-\sum_{\Gamma_j:\textrm{TRIM}}n_-(\Gamma_j)\quad (\textrm{mod}\ 4), \\
\nu_a&\equiv\sum_{\Gamma_j:\textrm{TRIM}\land k_a=\pi}n_-(\Gamma_j)\quad (\textrm{mod}\ 2)\ (a=1,2,3). 
\end{align}
We note that, for insulators, $\mu_1$ takes only the values 0 or 2. If $\mu_1=1$ or 3, there are Weyl points somewhere in $\vk$-space between the valence bands and the conduction bands, meaning that the system is not an insulator \cite{PhysRevB.83.245132,PhysRevB.85.165120}.

In this section, we show a direct correspondence between the bulk topological invariant $\mu_1$ and the existence of chiral hinge states. To this end, we introduce a cutting procedure along one direction, which we call $x_1$ axis, and study the 2D system after cutting the system along a plane $x_1=\text{const.}$, which corresponds to $\lambda=0$ in the cutting process. We show that in a system with an open boundary condition in one direction, when $\mu_1=2$, (i) the surface band structure is gapless or (ii) the surface band structure is gapped and the Chern number of the system is equal to 1 (mod 2). 
More precisely, under the assumption that the system is insulating both in the bulk and the surface, we show the following relation: 
\begin{align}
\sum_{\overline{\Gamma_j}}\tilde{n}_-(\overline{\Gamma_j})|_{\lambda=0}\equiv 
\frac{1}{2}(-\mu_1) \quad (\textrm{mod}\ 2), 
\label{Chern_mu_mod}
\end{align}
where $\tilde{n}_-(\overline{\Gamma_j})$ is the number of occupied states with odd parity at the 2D TRIM $\overline{\Gamma_j}$. 
Equation (\ref{Chern_mu_mod}) is a central result of this paper, which shows the direct relationship of the 3D indicator in the bulk and the 2D indicator with an open boundary. 
We note that, as is similar to the one-dimensional case, we take an inversion-symmetric unit cell which is commensurate with the open system.

\subsection{General proof}
\begin{figure}[t]
  \centerline{\includegraphics[width=8cm,clip]{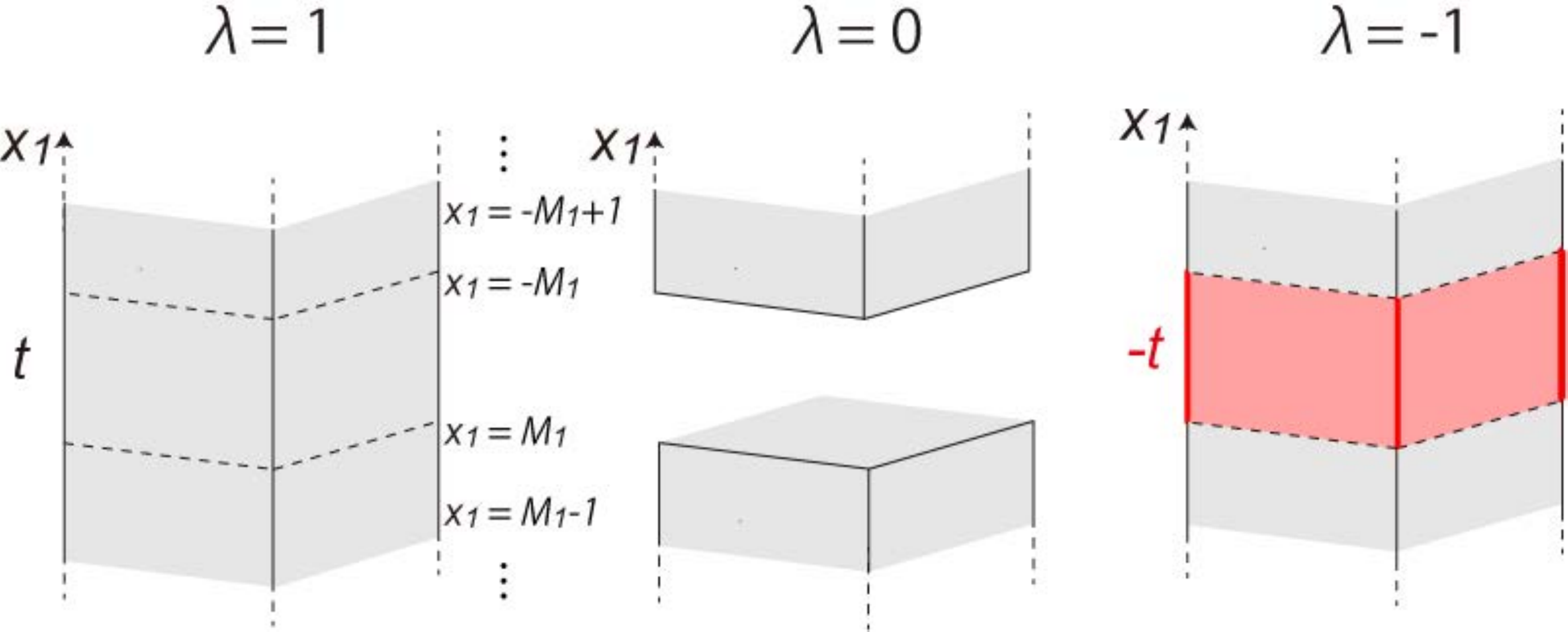}}
  \caption{(Color online) Conceptual figure of a three-dimensional inversion-symmetric insulator cut at the boundary between $x_1=M_1$ and $x_1=-M_1$. When $\lambda=1$, the system is periodic, and when $\lambda=0$ the system is an open finite system along the $x_1$ direction.  When $\lambda=-1$, this system is anti-periodic along the $x_1$ direction. 
}
    \label{3d_general_A}
\end{figure}

We set the system size along the $x_i$ axis to be $L_i\ (i=1,2,3)$ measured in units of $|\va_i|\ (i=1,2,3)$. 
For simplicity, let $L_1$ be an odd integer, $L_1=2M_1+1$ $(M_1:\text{integer})$, and let $L_2,L_3\to\infty$. The centers of the unit cells are located at $x_1=-M_1,-M_1+1,\cdots,M_1$, measured in the units of $|\va_1|$. 
We first start with a periodic system with $x_1=M_1$ and $x_1=-M_1$ being connected. We then replace the hopping amplitudes $t_j$ for all bonds that cross the boundary between $x_1=M_1$ and $x_1=-M_1$ by $\lambda t_j$, where $\lambda$ is real. 
Note that we impose a periodic boundary condition for $x_2$ and $x_3$ directions in the rest of this section. Figure \ref{3d_general_A} is a conceptual figure of a three-dimensional inversion-symmetric insulator cut at the boundary between $x_1=M_1$ and $x_1=-M_1$. For any real values of $\lambda$, inversion symmetry is always preserved.

By considering the degree of freedom in the $x_1$ direction as an internal degree of freedom, this system can be regarded as a two-dimensional system. Then we regard $\vk_{||}\equiv(k_2,k_3)^T$ as a wave vector in this effective 2D system, where $k_2$ and $k_3$ take the value $-\pi< k_i \leq \pi\ (i=2,3)$.  
$\vk_{||}$ is a TRIM if and only if $\vk_{||}=(0,0), (\pi,0), (0,\pi), (\pi,\pi)$, which we call $\overline{\Gamma}$, $\overline{Y}$, $\overline{Z}$, $\overline{T}$, respectively. 
Let $N_{\overline{\Gamma_j}}(\lambda)$ be the total number of occupied states at $\vk_{||}=\overline{\Gamma_j}$ for a given value of $\lambda$.

We show that $\tilde{n}_-(\overline{\Gamma_j})|_{\lambda=0}$ is evaluated from the knowledge of $n_-(\Gamma_j)$ and $N_{\overline{\Gamma_j}}|_{\lambda=0}$. 
We note that $\vk_{||}=\overline{\Gamma_j}$ represents a one-dimensional $\mathcal{P}$-invariant subspace in 3D $\vk$-space. In this subspace, $\tilde{n}_-(\overline{\Gamma_j})$ corresponds to $N_-$ in the one-dimensional case discussed in Sec.~I\hspace{-.1em}I. 
Therefore, from Eqs.~(\ref{Nn0}), (\ref{Nnpi}) and (\ref{nuopen_nubulk2}), the following equations hold:
\begin{align}
\tilde{n}_-(\overline{\Gamma_j})|_{\lambda=1}&=\frac{(L_1-1)\nu}{2}+n_{-}(0,\overline{\Gamma_j}), \label{NnG0}
\\
\tilde{n}_-(\overline{\Gamma_j})|_{\lambda=-1}&=\frac{(L_1-1)\nu}{2}+n_{-}(\pi,\overline{\Gamma_j}), \label{NnGpi}
\\
[\tilde{n}_-(\overline{\Gamma_j})]^{\lambda=0}_{\lambda=1}
&=\frac{1}{2}[N_{\overline{\Gamma_j}}]^{\lambda=0}_{\lambda=1}
+\frac{n_-(\pi,\overline{\Gamma_j})-n_-(0,\overline{\Gamma_j})}{2}, 
\label{nuopen_nubulk_g}
\end{align}
where $(0,\overline{\Gamma_j})$ and $(\pi,\overline{\Gamma_j})$ are 3D TRIM. 
By adding Eq.~(\ref{NnG0}) and Eq.~(\ref{nuopen_nubulk_g}), we get the following equation: 
\begin{align}
\tilde{n}_-(\overline{\Gamma_j})|_{\lambda=0}
=\frac{N_{\overline{\Gamma_j}}|_{\lambda=0}-\nu}{2}
&+\frac{n_-(0,\overline{\Gamma_j})+n_-(\pi,\overline{\Gamma_j})}{2}, 
\label{ntilde_gammaj_lambda0}
\end{align}
where we use the relation $N_{\overline{\Gamma_j}}|_{\lambda=1}=L_1\nu$. 
By summing Eq.~(\ref{ntilde_gammaj_lambda0}) over $\overline{\Gamma_j}$, we obtain the following equation: 
\begin{align}
\sum_{\overline{\Gamma_j}}\tilde{n}_-(\overline{\Gamma_j})|_{\lambda=0}
=&\frac{1}{2}\sum_{\overline{\Gamma_j}}N_{\overline{\Gamma_j}}|_{\lambda=0}-2\nu \notag \\
&\quad+\frac{1}{2}\sum_{\Gamma_j}n_-(\Gamma_j). 
\end{align}
Next, let us assume that the system at $\lambda=0$ is also insulating, which means that the surface is also insulating. Then the values of $N_{\overline{\Gamma_j}}|_{\lambda=0}$ at the four TRIM are equal, i.e. $N_{\overline{\Gamma_j}}|_{\lambda=0}=N_{\textrm{open}}$. 
Then, we get the following equation: 
\begin{align}
\sum_{\overline{\Gamma_j}}\tilde{n}_-(\overline{\Gamma_j})|_{\lambda=0}
&=2N_{\textrm{open}}-2\nu+\frac{1}{2}(-\mu_1). \label{chern_mu}
\end{align}
Then we note that the l.h.s. of Eq.~(\ref{chern_mu}) is equal to the parity of the Chern number at $\lambda=0$. It is because, in a two-dimensional centrosymmetric system, $(-1)^{\textrm{Ch}}$ is equal to the product of the $\mathcal{P}_{\text{2D}}$ eigenvalues at TRIM, where $\mathcal{P}_{\text{2D}}$ is the inversion operator of the two-dimensional system, as shown in the Sec.~I\hspace{-.1em}I\hspace{-.1em}I and in some previous works\cite{PhysRevB.83.245132,PhysRevB.85.165120,PhysRevB.86.115112}. 
Equation (\ref{Chern_mu_mod}) is derived by taking modulo $2$ in Eq.~(\ref{chern_mu}).

\begin{figure}[t]
  \centerline{\includegraphics[width=8cm,clip]{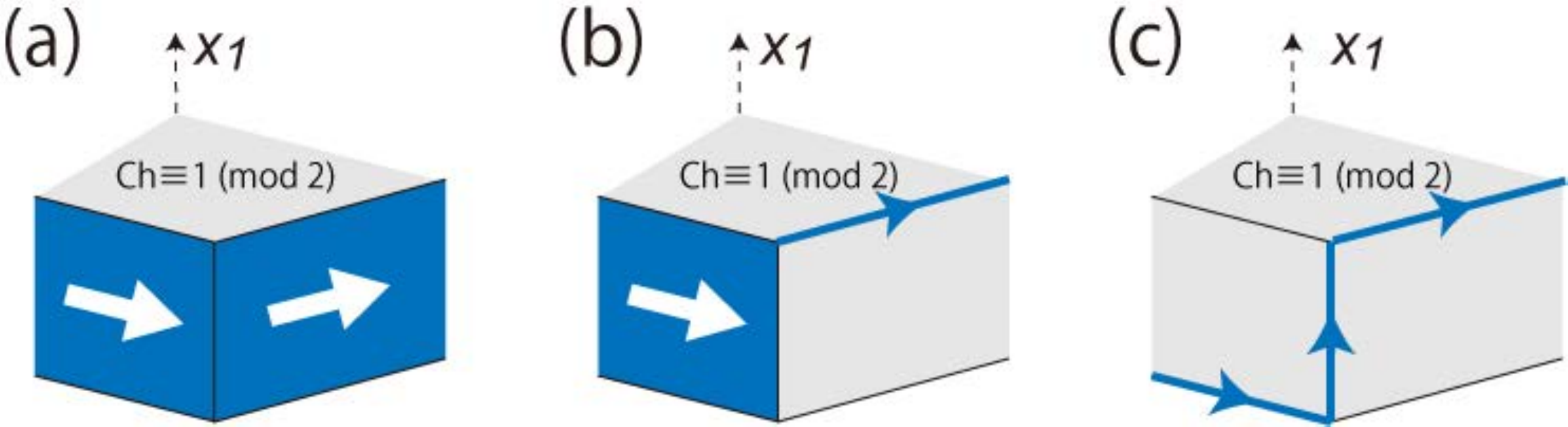}}
  \caption{(Color online) Conceptual figure of positions of gapless surface states and hinge states for $\mu_1=2$. In this case, when the system is cut along the $x_2$-$x_3$ plane, the resulting 2D system has the Chern number equal to 1 (mod 2), meaning that the 2D system has gapless chiral edge states. These chiral edge states as a 2D system can be either surface states or hinge states. In (a), the gapless chiral states are surface states at the side surfaces. In (b), the gapless chiral states are composed of surfaces states and hinge states. In (c), they are hinge states. 
}
    \label{3d_general_B}
\end{figure}

In particular, when $\mu_1=2$, we conclude that $(\textrm{Chern})|_{\lambda=0}=1$ (mod 2). It means that if we make the system to be open also along $x_2$-, and $x_3$-directions, this system supports chiral edge modes with the number of chiral modes being an odd number. These chiral modes can be on surfaces or on hinges, as shown in Fig.~\ref{3d_general_B}(a-c). If we further assume that all the surfaces are gapped, these chiral modes are localized on the hinges.

\subsection{Illustrative examples}
\begin{figure}[t]
  \centerline{\includegraphics[width=8cm,clip]{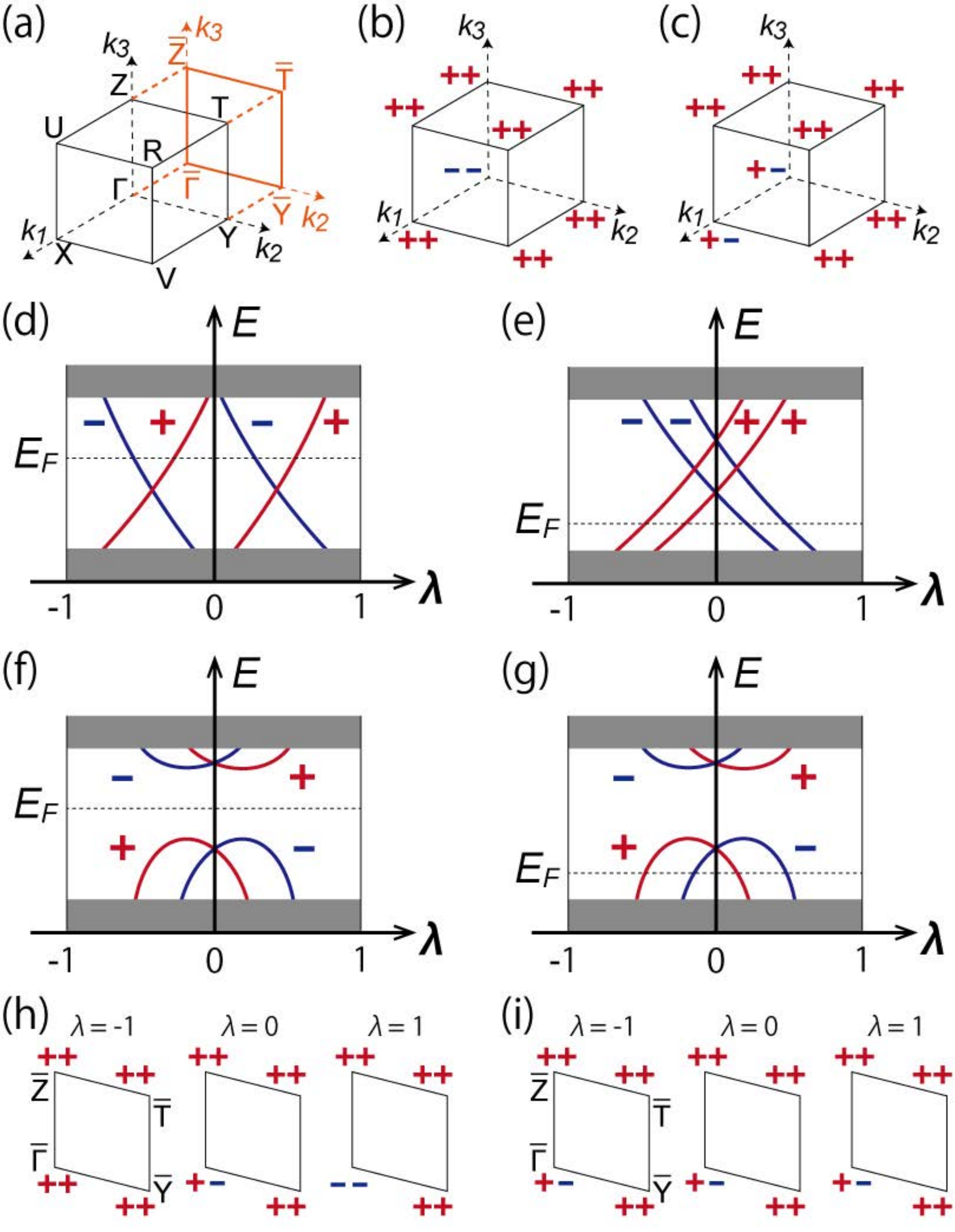}}
  \caption{(Color online) Illustrative examples of $\mu_1=2$ topological insulators. (a) $1/8$ of the 3D Brillouin zone and the symmetry labels shown in black, and $1/4$ of the 2D Brillouin zone and the symmetry labels shown in orange. (b-c) Two examples of inversion parities at TRIM for (b) $\nu_a=0$ (HOTI) and (c) $\nu_1=1$, $\nu_2=\nu_3=0$ (weak Chern insulator). 
(d-g) Four representative examples of the energy spectrum on $\lambda$. Red and blue lines represent bound states with even and odd parity, respectively. (f) and (g) differ only by the Fermi energy. (h-i) Two examples of inversion parities at 2D TRIM corresponding to (b) and (c) respectively. In every case, in addition to the state whose parity is shown in (h) and (i), there are $\frac{(L_1-1)\nu}{2}$ states with even parity and $\frac{(L_1-1)\nu}{2}$ states with odd parity, and they are not shown in the figure for simplicity.  When $\lambda=0$, both examples have one state with odd parity at $\overline{\Gamma}$. 
}
    \label{parity_exchange_3d}
\end{figure}

In this subsection, we consider two typical examples and demonstrate the validity of the formula (\ref{Chern_mu_mod}) holds: (A) $\mu_1=2,\nu_a=0$ (HOTI) and (B) $\mu_1=2,\nu_1=1, \nu_2=\nu_3=0$ (weak Chern insulator). In the rest of this subsection, we assume that the surface band structure, i.e. the band structure at $\lambda=0$, is gapped. Since $\mu_1=2$ in both cases, from the formula (\ref{Chern_mu_mod}), both systems have a nontrivial Chern number when $\lambda=0$. However, due to the difference of the weak topological number, $\nu_1$, their topological nature is completely different. Here, we illustrate the spectral flow for the change of $\lambda$, and we confirm the result by a model calculation. We find that the spectral flows of (A) and (B) are completely different. The difference comes from the difference of $\nu_1$.

\subsubsection{$\mu_1=2,\nu_a=0\ (a=1,2,3)$}
One of the simplest examples to realize $\mu_1=2,\nu_a=0\ (a=1,2,3)$ is $n_-(\Gamma)=2$, $n_-(\Gamma_j)=0\ (\Gamma_j\neq\Gamma)$ and $\nu=2$. 
The bulk parity corresponding to this example is illustrated in Fig.~\ref{parity_exchange_3d} (b). 
We note that, since $\nu=2$ and $L_1-1=2M_1$, from Eq.~(\ref{NnG0}) and (\ref{NnGpi}), $\tilde{n}_{-}(\overline{\Gamma_j})|_{\lambda=1}=2M_1 + n_{-}(0,\overline{\Gamma_j})$, and $\tilde{n}_{-}(\overline{\Gamma_j})|_{\lambda=-1}=2M_1+ n_{-}(\pi,\overline{\Gamma_j})$. 
Then following relations hold: 
\begin{align}
\tilde{n}_{-}(\overline{\Gamma})|_{\lambda=1}&=2M_1+2, \quad 
\tilde{n}_{-}(\overline{\Gamma})|_{\lambda=-1}=2M_1, 
\label{n_Gamma_1-1}
\\
\tilde{n}_{-}(\overline{Y})|_{\lambda=\pm1}&=\tilde{n}_{-}(\overline{Z})|_{\lambda=\pm1}=\tilde{n}_{-}(\overline{T})|_{\lambda=\pm1}=2M_1. 
\end{align}

Next we calculate $\tilde{n}_-(\overline{\Gamma_j})|_{\lambda=0}$ for TRIM $\overline{\Gamma_j}$. 
First, let us consider the case of $\vk_{||}=\overline{\Gamma}$. 
From Eq.~(\ref{n_Gamma_1-1}), when $\lambda$ is decreased from $1$ to $-1$, two odd-parity states cross the Fermi energy $E_F$ from below, and two even parity states cross $E_F$ from above. Furthermore, as we show in Appendix A, the energy spectrum is symmetric under the flipping of the sign of $\lambda$, and the bound states $\ket{\psi_l(\lambda)}$ and $\ket{\psi_l(-\lambda)}$ have opposite parities.
Figure \ref{parity_exchange_3d}(d-e) shows two representative examples of the energy spectrum which satisfy these requirements. 
The number of states below the Fermi energy $E_F$ at $\lambda=0$, $N_{\overline{\Gamma}}|_{\lambda=0}$, is equal to that of bulk in (d) and two less than that of bulk in (e). 
The differences in the value of $\tilde{n}_{-}(\overline{\Gamma})$ between $\lambda=0$ and $\lambda=1$ are $[\tilde{n}_{-}(\overline{\Gamma})]^{\lambda=0}_{\lambda=1}=-1$ in Fig.~\ref{parity_exchange_3d}(d) and $[\tilde{n}_{-}(\overline{\Gamma})]^{\lambda=0}_{\lambda=1}=-2$ in (e). 
Therefore, $\tilde{n}_{-}(\overline{\Gamma})|_{\lambda=0}\equiv1$ (mod 2) in Fig.~\ref{parity_exchange_3d}(d), and $\tilde{n}_{-}(\overline{\Gamma})|_{\lambda=0}\equiv0$ (mod 2) in (e).

Next, let us consider the case of $\vk_{||}=\overline{Y}$. 
We note that the cases of $\vk_{||}=\overline{Z},\overline{T}$ are the same as the case of $\vk_{||}=\overline{Y}$. 
We consider the energy spectrum at $\overline{Y}$ in changing $\lambda$ from $1$ to $-1$. 
Since $\tilde{n}_{-}(\overline{Y})|_{\lambda=\pm1}=2M_1$, the number of states with even (odd) parity which cross $E_F$ from above is equal to that from below through the change of $\lambda$ from $\lambda=1$ to $\lambda=-1$. 
Figure \ref{parity_exchange_3d}(f-g) shows two representative examples of energy spectrum which satisfy this requirement.
In Fig.~\ref{parity_exchange_3d}(f), $N_{\overline{Y}}|_{\lambda=0}$ is equal to that of bulk, and in Fig.~\ref{parity_exchange_3d}(g), $N_{\overline{Y}}|_{\lambda=0}$ is two less than that of bulk. 
The differences in the value of $\tilde{n}_{-}(\overline{Y})$ between $\lambda=0$ and $\lambda=1$ are $[\tilde{n}_{-}(\overline{Y})]^{\lambda=0}_{\lambda=1}=0$ in Fig.~\ref{parity_exchange_3d}(f) and $[\tilde{n}_{-}(\overline{Y})]^{\lambda=0}_{\lambda=1}=-1$ in Fig.~\ref{parity_exchange_3d}(g). 
Therefore, 
$\tilde{n}_{-}(\overline{Y})|_{\lambda=0}\equiv0$ (mod 2) in Fig.~\ref{parity_exchange_3d}(f), and $\tilde{n}_{-}(\overline{Y})|_{\lambda=0}\equiv1$ (mod 2) in (g).

Finally, we combine the results at $\overline{\Gamma}$, $\overline{Y}$, $\overline{Z}$ and $\overline{T}$. 
First of all, from the assumption that the bulk is insulating, 
the number of bulk occupied states is equal for all the 2D TRIM $\overline{\Gamma_j}$: 
\begin{align}
N|_{\lambda=1}
\equiv
N_{\overline{\Gamma}}|_{\lambda=1}
=N_{\overline{Y}}|_{\lambda=1}
=N_{\overline{Z}}|_{\lambda=1}
=N_{\overline{T}}|_{\lambda=1}. 
\end{align}
Secondly, from the assumption that the surface is insulating, the following equation holds: 
\begin{align}
N|_{\lambda=0}
\equiv
N_{\overline{\Gamma}}|_{\lambda=0}
=N_{\overline{Y}}|_{\lambda=0}
=N_{\overline{Z}}|_{\lambda=0}
=N_{\overline{T}}|_{\lambda=0}. 
\end{align}
Hence, there are two possibilities, depending on the difference between $N|_{\lambda=1}$ and $N|_{\lambda=0}$ being $0$ or $2$ modulo $4$. 
This constrains combinations of evolutions of states at $\overline{\Gamma}$, $\overline{Y}$, $\overline{Z}$ and $\overline{T}$ upon a change of $\lambda$. 
First, when $N|_{\lambda=1}=N|_{\lambda=0}$, the spectrum at $\overline{\Gamma}$ is like Fig.~\ref{parity_exchange_3d}(d), while that at $\overline{Y}$, $\overline{Z}$ and $\overline{T}$ are Fig.~\ref{parity_exchange_3d}(f). 
Therefore, $\tilde{n}_{-}(\overline{\Gamma})|_{\lambda=0}\equiv1$ and $\tilde{n}_{-}(\overline{Y})|_{\lambda=0}\equiv\tilde{n}_{-}(\overline{Z})|_{\lambda=0}\equiv\tilde{n}_{-}(\overline{T})|_{\lambda=0}\equiv0$ (mod 2), and the sum of the number of states with odd parity at 2D TRIM $\overline{\Gamma_j}$ is 1 (mod 2). 
Figure \ref{parity_exchange_3d}(h) shows inversion parities at $\lambda=-1,0$ and $1$, corresponding to Fig.~\ref{parity_exchange_3d}(d),(f). 
Second, when $N|_{\lambda=1}-N|_{\lambda=0}\equiv2$ (mod 4), the spectrum at $\overline{\Gamma}$ is Fig.~\ref{parity_exchange_3d}(e) while that at $\overline{Y}$, $\overline{Z}$ and $\overline{T}$ are Fig.~\ref{parity_exchange_3d}(g). 
Therefore, $\tilde{n}_{-}(\overline{\Gamma})|_{\lambda=0}\equiv0$ and $\tilde{n}_{-}(\overline{Y})|_{\lambda=0}\equiv\tilde{n}_{-}(\overline{Z})|_{\lambda=0}\equiv\tilde{n}_{-}(\overline{T})|_{\lambda=0}\equiv1$, and their sum is $3\equiv1$ (mod 2). 
To summarize, in either case, $\sum_{\overline{\Gamma_j}}\tilde{n}_{-}(\overline{\Gamma_j})|_{\lambda=0}=1$ (mod 2), which means that the Chern number of the system is 1 (mod 2).

We confirm this by a model calculation. 
We use the following tight-binding Hamiltonian:
\begin{align}
\mathcal{H}(\vk)
&=-\Big{(}m-c\sum_{j=x,y,z}\cos k_j\Big{)}\tauz \notag \\
&\quad-t\sum_{j=x,y,z}\sigma_j \taux\sin k_j  \notag \\
&\quad+\vB\cdot\mathbb{\sigma}+A\tauz\sigmaz,
\label{model_HOTI}
\end{align}
where $m=4, c=2, t=1$, $\vB=(0.3, 0.3, 0.3)$ and $A=0.3$. 
This Hamiltonian is symmetric under inversion operation $\mathcal{P}=\tauz$, and the inversion parities at TRIM are shown in Fig.~\ref{parity_exchange_3d}(b). If $\vB=\mathbf{0}$ and $A=0$, the model (\ref{model_HOTI}) describes a conventional topological insulator protected by time-reversal symmetry\cite{PhysRevB.98.205129,PhysRevB.78.195424}. $\vB$ is considered as a uniform magnetic field, which breaks time-reversal symmetry. $A$ is considered as an orbital-dependent magnetic field which depends on the orbital degrees of freedom $\tau$. 
The Fermi energy is set to be $E_F=0$. The energy spectrum on $\lambda$ at $\overline{\Gamma}$ and $\overline{Y}$ is shown in Fig.~\ref{numerical} (a) and (b) respectively. 
As expected from the theoretical calculation, the energy spectrum at $\overline{\Gamma}$ has crossing points as shown in Fig.~\ref{numerical} (a), and there is no crossing point at $\overline{Y}$ as shown in Fig.~\ref{numerical} (b). This corresponds to the case with Fig.~\ref{parity_exchange_3d}(d) and (f). 
These results support our theoretical calculation.

\subsubsection{$\mu_1=2,\nu_1=1, \nu_2=\nu_3=0$}
As an example to realize $\mu_1=2,\nu_1=1, \nu_2=\nu_3=0$, here we take $n_-(\Gamma)=n_-(X)=1$, $n_-(\Gamma_j)=0\ (\Gamma_j\neq\Gamma,X)$ and $\nu=2$. 
The bulk parity corresponding to this example is illustrated in Fig.~\ref{parity_exchange_3d} (c). 

First, we note that the cases of $\vk_{||}=\overline{Y}$, $\overline{Z}$ and $\overline{T}$ are the same as the case of $\vk_{||}=\overline{Y}$ in the previous example in Sec.~I\hspace{-.1em}V B1. Therefore, for $\overline{\Gamma_j}=\overline{Y}$, $\overline{Z}$ and $\overline{T}$, we get $\tilde{n}_{-}(\overline{\Gamma_j})|_{\lambda=0}\equiv0$ (mod 2) in Fig.~\ref{parity_exchange_3d}(f), and $\tilde{n}_{-}(\overline{\Gamma_j})|_{\lambda=0}\equiv1$ (mod 2) in (g).

Next, let us consider the case of $\vk_{||}=\overline{\Gamma}$. 
Since $\nu=2$ and $L_1-1=2M_1$, from Eq.~(\ref{NnG0}) and (\ref{NnGpi}), we get $\tilde{n}_{-}(\overline{\Gamma})|_{\lambda=1}=2M_1+n_{-}(\Gamma)=2M_1+1$, and $\tilde{n}_{-}(\overline{\Gamma})|_{\lambda=-1}=2M_1+n_{-}(X)=2M_1+1$. 
Therefore, through a change from $\lambda=1$ to $\lambda=-1$, the number of states with even (odd) parity which cross $E_F$ from above is equal to that from below as a whole. Two examples of the possible energy spectra are shown in Fig.~\ref{parity_exchange_3d}(f) and (g). 
We get $\tilde{n}_{-}(\overline{\Gamma})|_{\lambda=0}\equiv1$ (mod 2) in Fig.~\ref{parity_exchange_3d}(f), and $\tilde{n}_{-}(\overline{\Gamma})|_{\lambda=0}\equiv0$ (mod 2) in (g).

Finally, we combine the results at $\overline{\Gamma}$, $\overline{Y}$, $\overline{Z}$ and $\overline{T}$. First of all, from the assumption that both the bulk and the surface is insulating, $N_{\overline{\Gamma_j}}|_{\lambda=0}$ and  $N_{\overline{\Gamma_j}}|_{\lambda=1}$ do not depend on $\overline{\Gamma_j}$.
Therefore, in Fig.~\ref{parity_exchange_3d}, within the two cases (f) and (g), the energy spectra at the four 2D TRIM $\overline{\Gamma_j}$ ($\overline{\Gamma_j}=\overline{\Gamma}, \overline{Y}, \overline{Z}, \overline{T}$) are the same,
because $N_{\lambda=1}-N_{\lambda=0}$ is equal to 0 (mod 4) in Fig.~\ref{parity_exchange_3d}(f) and 2 (mod 4) in Fig.~\ref{parity_exchange_3d}(g). 
In Fig.~\ref{parity_exchange_3d}(f), since $\tilde{n}_{-}(\overline{\Gamma})|_{\lambda=0}\equiv1$ and $\tilde{n}_{-}(\overline{Y})|_{\lambda=0}\equiv\tilde{n}_{-}(\overline{Z})|_{\lambda=0}\equiv\tilde{n}_{-}(\overline{T})|_{\lambda=0}\equiv0$, 
their sum is 1 (mod 2). 
Figure \ref{parity_exchange_3d}(i) shows inversion parities at $\lambda=-1,0$ and $1$, corresponding to Fig.~\ref{parity_exchange_3d}(f). 
In Fig.~\ref{parity_exchange_3d}(g), since $\tilde{n}_{-}(\overline{\Gamma})|_{\lambda=0}\equiv0$ and $\tilde{n}_{-}(\overline{Y})|_{\lambda=0}\equiv\tilde{n}_{-}(\overline{Z})|_{\lambda=0}\equiv\tilde{n}_{-}(\overline{T})|_{\lambda=0}\equiv1$, their sum is $3\equiv1$ (mod 2). 
In either case, $\sum_{\overline{\Gamma_j}}\tilde{n}_{-}(\overline{\Gamma_j})|_{\lambda=0}=1$ (mod 2), which means that the Chern number of the system is 1 (mod 2).

\begin{figure}[t]
  \centerline{\includegraphics[width=8.5cm,clip]{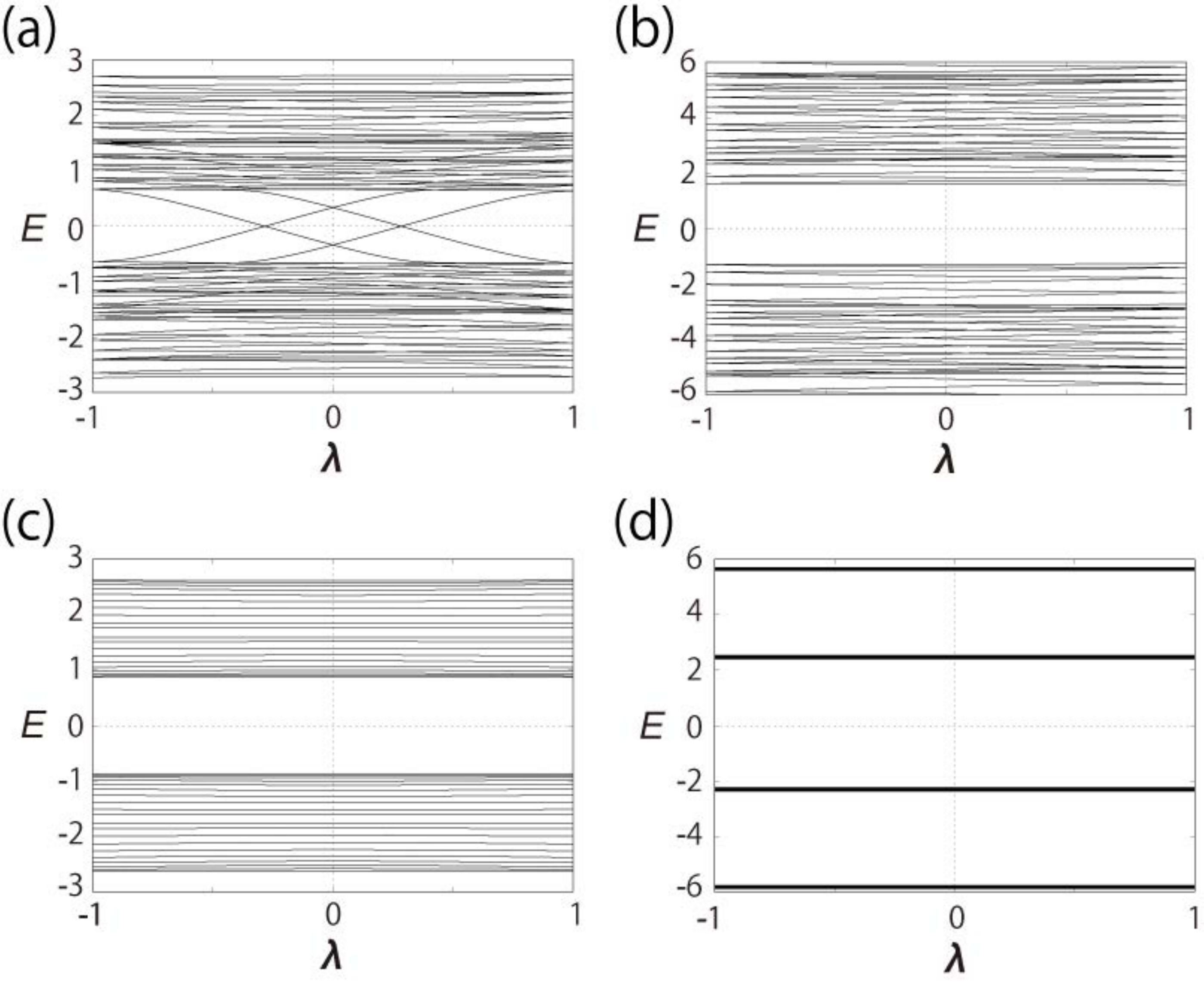}}
  \caption{Model calculations for two examples of $\mu_1=2$ topological insulators, (a-b) a HOTI and (c-d) a weak Chern insulator corresponding to cases (A) and (B) in Sec.~I\hspace{-.1em}V B. 
(a-b) Energy spectrum of the model (\ref{model_HOTI}) for a HOTI as a function of $\lambda$ at (a) $\overline{\Gamma}$ and (b) $\overline{Y}$. 
(c-d) Energy spectrum of the model (\ref{model_WTI}) for a weak Chern insulator as a function of $\lambda$ at (c) $\overline{\Gamma}$ and (d) $\overline{Y}$.
}\label{numerical}
\end{figure}

We confirm this by model calculation. 
We use the following tight-binding Hamiltonian:
\begin{align}
\mathcal{H}(\vk)
&=-\Big{(}m-c\sum_{j=y,z}\cos k_j\Big{)}\tauz \notag \\
&\quad-t\sum_{j=x,y,z}\sigma_j \taux\sin k_j  \notag \\
&\quad+\vB\cdot\mathbb{\sigma}+A\tauz\sigmaz,
\label{model_WTI}
\end{align}
where $m=4, c=2, t=1$, $\vB=(1.5, 0.5, 0.5)$ and $A=0.3$. 
This Hamiltonian is symmetric under inversion operation $\mathcal{P}=\tauz$, and the inversion parities at TRIM are shown in Fig.~\ref{parity_exchange_3d}(c). The main difference between the models (\ref{model_HOTI}) and (\ref{model_WTI}) is presence/absence of the term $\tauz\cos k_x$. 
The energy spectrum on $\lambda$ at $\overline{\Gamma}$ and $\overline{Y}$ is shown in Fig.~\ref{numerical} (c) and (d) respectively. 
As expected from the theoretical calculation, the energy spectrum at $\overline{\Gamma}$ and $\overline{Y}$ have no crossing point as shown in Fig.~\ref{numerical} (c) and (d) respectively. 
From these calculations, in the case with $(\mu_1,\nu_1)=(2,1)$, we conclude that the Chern number perpendicular to the $x_1$-axis is always 1: $(\text{Chern})|_{\lambda=1}=(\text{Chern})|_{\lambda=0}=1$ (mod 2). This is different from the previous example with $(\mu_1,\nu_1)=(2,0)$. In that case, the Chern number depends on $\lambda$: $(\text{Chern})|_{\lambda=1}=0$ (mod 2) and $(\text{Chern})|_{\lambda=0}=1$ (mod 2).

\section{Bulk-edge and Bulk-hinge correspondence in class AII}
In this section, we extend the results of the previous sections to spinful systems with time-reversal symmetry, i.e. class AII systems.
In class AII systems, due to the time-reversal symmetry, all energy eigenstates are Kramers-degenerate. Therefore, the number of occupied states $N$ should be an even number. 
Moreover, since the inversion operator $\mathcal{P}$ commutes with the time-reversal operator $\Theta$, all the inversion eigenstate $\ket{\psi}$ and its Kramers partner $\Theta\ket{\psi}$ have the same parity.  This means that, the number of occupied states with odd parity is always an even number. Therefore, $N_-$ and $n_-(\Gamma_j)$ should be even. 
Here, we define $M$, $M_-$ and $m_-(\Gamma_j)$ as halves of $N$, $N_-$ and $n_-(\Gamma_j)$, respectively, i.e., 
\begin{align}
M=\frac{N}{2}, 
\quad
M_-=\frac{N_-}{2}, 
\quad
m_-(\Gamma_j)=\frac{n_-(\Gamma_j)
}{2}. 
\end{align}
By replacing $N$, $N_-$ and $n_-$ in the previous sections with $M$, $M_-$ and $m_-$ respectively, we obtain an extension of our theory to class AII systems. 
We note that $M$ is equal to the total number of occupied Kramers pairs of states. Likewise, $M_-$ is equal to the number of occupied Kramers pairs of states with odd parity, and $m_-(\Gamma_j)$ is the number of occupied Kramers pairs of states with odd parity at a TRIM $\Gamma_j$. 
We also note that one- and two-dimensional cases are already discussed in the previous study\cite{PhysRevB.78.045426}. 
However, for convenience of the readers, we rewrite these discussions in the previous study in our notation. 

\subsection{1D}
By replacing $N_-$ and $n_-$ in Eq.~(\ref{nuopen_nubulk2}) with $M_-$ and $m_-$, respectively, we obtain the following relation: 
\begin{align}
M_{\textrm{open}}
&=M_{\textrm{bulk}}+m_-(0)-m_-(\pi)+2[M_{-}]^{\lambda=0}_{\lambda=1},  
\label{mopen_mbulk2}
\\
&\equiv M_{\textrm{bulk}}+m_-(0)-m_-(\pi) \quad(\text{mod}\ 2).  
\label{mopen_mbulk2mod}
\end{align}
This means that the number of occupied Kramers pairs of states at $\lambda=0$ and $\lambda=1$ differ by 1 (mod 2) if $m_-(0)-m_-(\pi)\equiv1$ (mod 2). 

\subsection{2D}
By replacing $N$ and $n_-$ in Eq.~(\ref{nu_chern}) with $M$ and $m_-$ respectively, we obtain the following relation: 
\begin{align}
M_0|_{\lambda=0}-M_{\pi}|_{\lambda=0}
&\equiv \sum_{\Gamma_j\in\textrm{TRIM}} m_-(\Gamma_j) \quad (\textrm{mod 2}). 
\label{nu_chern_M}
\end{align}
This means that if the r.h.s. of Eq.~(\ref{nu_chern_M}) is 1 mod 2, 
the number of occupied Kramers pairs of states for $\lambda=0$ at $k_{||}=0$ is different from that at $k_{||}=\pi$\cite{PhysRevB.78.045426}. 
In order to compensate the difference, there should be an odd number of  helical edge modes between $k_{||}=0$ and $k_{||}=\pi$. 
We note that this result is consistent with the well-known Fu-Kane formula\cite{PhysRevB.76.045302}, which shows that the r.h.s. of Eq.~(\ref{nu_chern_M}) is equal to the $Z_2$ topological invariant of the system.

\subsection{3D}
By replacing $N$ and $n_-$ in Eq.~(\ref{chern_mu}) with $M$ and $m_-$ respectively, we obtain the following relation: 
\begin{align}
\sum_{\overline{\Gamma_j}}\tilde{m}_-(\overline{\Gamma_j})|_{\lambda=0}
&=2M|_{\lambda=0}-\nu+\frac{1}{2}(-\kappa_1) \notag \\ 
&\equiv \frac{1}{2}(-\kappa_1) \quad (\textrm{mod}\ 2). \label{Z2_kappa}
\end{align}
Here, $\kappa_1$ is defined as follows\cite{PhysRevB.98.115150}: 
\begin{align}
\kappa_1
&=\sum_{\Gamma_j:\textrm{TRIM}}(n_+(\Gamma_j)-n_-(\Gamma_j))/4
\notag \\
&\equiv-\sum_{\Gamma_j:\textrm{TRIM}}n_-(\Gamma_j)/2
\quad (\text{mod}\ 4)
\notag \\
&\equiv-\sum_{\Gamma_j:\textrm{TRIM}}m_-(\Gamma_j)
\quad (\text{mod}\ 4). 
\end{align}
The l.h.s of Eq.(\ref{Z2_kappa}) is equal to the $Z_2$ topological invariant of the system in a slab geometry\cite{PhysRevB.76.045302}. Therefore, if $\kappa_1=2$ (mod 4), the system in a slab geometry is a quantum spin Hall insulator. 
It means that if we make the system to be open also along $x_2$-, and $x_3$-directions, this system supports helical edge modes with the number of helical modes being an odd number. 
As in the previous section, these ``edge'' modes can be on surfaces or on hinges. 
If we further assume that all the surfaces are gapped, these helical edge modes are at the hinges. 
This gives a proof of the bulk-hinge correspondence in inversion-symmetric 3D class AII systems\cite{PhysRevX.8.031070,PhysRevB.97.205136,PhysRevB.98.081110}.

\section{Conclusion}
In the present paper, we studied the bulk-edge and bulk-hinge correspondences in inversion-symmetric insulators. 
We used a cutting procedure, and study the spectral flow in the cutting process.
In one- and two-dimensional centrosymmetric systems, we showed that the proof of the bulk-edge correspondence is simplified by introducing a cutting procedure. 
For a three-dimensional centrosymmetric system, we proved the bulk-hinge correspondence by considering the spectral flow in the cutting process. 
Unlike the previous approach for the explanation of the bulk-hinge correspondence, our proof is applicable to more general tight-binding models with inversion symmetry.
We also confirmed this by model calculations, and showed that the spectral flow is consistent with the theoretical calculation.

One of the advantages of our method is its generality. 
Our method is applicable to any tight-binding models with inversion symmetry, as long as the unit cell is taken to be invariant under the inversion symmetry. 
In the main text, we have only considered systems with an odd value of $L$, the number of unit cells in the system. However, as shown in Appendices B and C, our proof can also be applied to systems with even $L$. 
Moreover, in the main text, we only consider the simple boundary condition. However, as shown in Appendix D, our approach is also applicable to other boundary conditions, as long as $L$ is sufficiently large. 

\begin{figure}[t]
  \centerline{\includegraphics[width=7cm,clip]{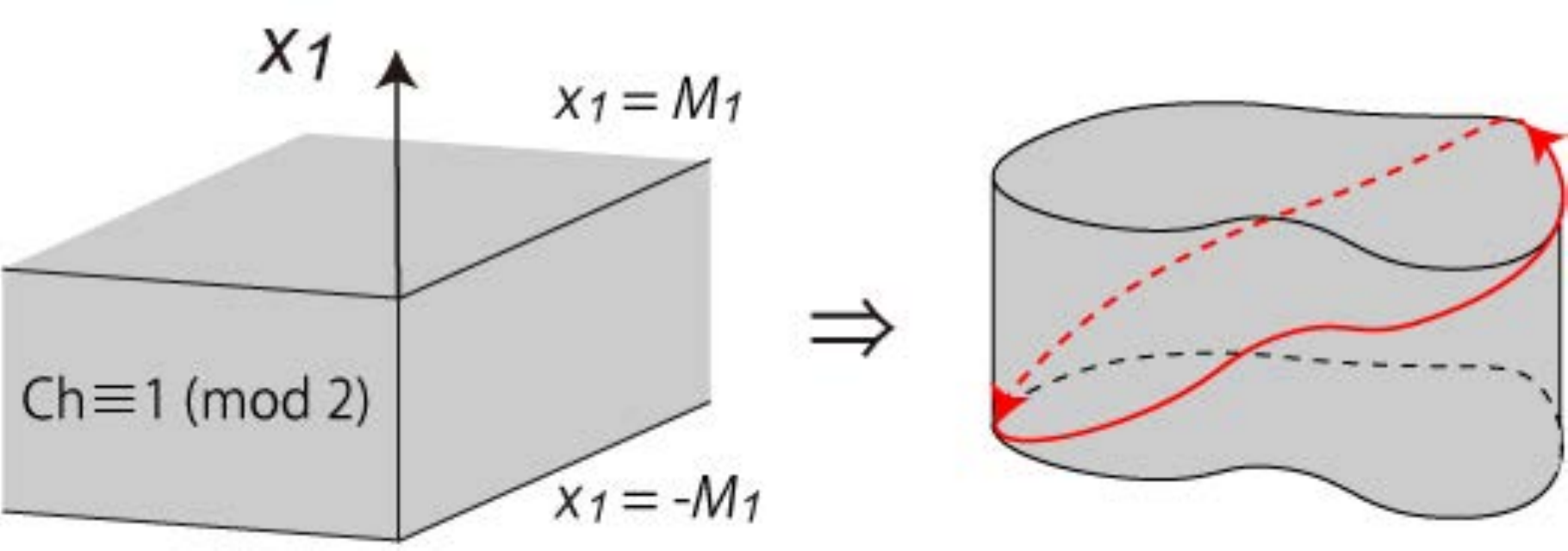}}
  \caption{(Color online) Conceptual figure of the boundary geometry and the gapless 1D states of the 3D inversion-symmetric HOTI. Red lines represent the gapless hinge states. As long as an infinite slab with surfaces at $x_1=\pm M_1$ preserves the inversion symmetry, gapless hinge states appear on the side surfaces.
}\label{boundary_geometry}
\end{figure}

Our result brings about a new perspective to an interplay between topology and system geometry of the HOTI. In the present paper, we showed that, when we impose an open boundary condition in one direction, a 3D inversion-symmetric HOTI in class A (class AII) becomes a 2D quantum Hall insulator (2D quantum spin Hall insulator). Since the edge modes of these 2D topological insulators appear irrespective of the shape of the boundary, the hinge states of these 3D HOTI always appear irrespective of the boundary shape of the other two directions, $x_2$ and $x_3$, as shown in Fig.~\ref{boundary_geometry}. This means that even if the crystal shape of the HOTI does not respect the inversion symmetry, the ``hinge'' states remain gapless as long as the crystal shape is wide enough in the $x_2$-$x_3$ direction (compared with the localization length of the hinge states) as shown in Fig.~\ref{boundary_geometry}.

\begin{acknowledgments}
R.T. thanks Motoaki Hirayama for useful comments and discussions. 
This work was supported by JSPS KAKENHI Grant Numbers JP18J23289 and JP18H03678.
\end{acknowledgments}

\appendix{}
\section{Unitary operator $U_x$ and approximation of $\mathcal{H}(-\lambda)$}

In this appendix we show two results which are used in Sec.~I\hspace{-.1em}I: (i) the Hamiltonian with the anti-periodic boundary condition is unitary equivalent to that with the periodic boundary condition with a shifted Bloch wave vector, and (ii) for sufficiently large $L$, the bound states $\ket{\psi_l(\lambda)}$ and $\ket{\psi_l(-\lambda)}$ have opposite parities. 
Both are derived from the fact that $\mathcal{H}(-\lambda)$ is well-approximated by $U_x\mathcal{H}(\lambda)U_x^{\dagger}$, where 
$U_x=\exp[i\pi\hat{x}/L]$ is a phase twist operator along $x$.

First, we see how $U_x\mathcal{H}(\lambda)U_x^{\dagger}$ is related to $\mathcal{H}(-\lambda)$. 
We consider a one-dimensional periodic system with the coordinate $x$. Let the system size in $x$-direction be $L=2M+1$ with an integer $M$ measured in the unit of the lattice constant $|\va|$. 
We introduce an open boundary via cutting procedure. We replace the hopping amplitudes $t_j$ for all the bonds that cross the boundary between $x=-M$ and $x=M$ by $\lambda t_j$. 
For simplicity, at first, we only consider the nearest neighbor hopping. 
Then the Hamiltonian is expressed as follows: 
\begin{align}
\mathcal{H}(\lambda)
&=
\begin{pmatrix}
H_0 & H_1^{\dagger} &  & \lambda H_1\\
H_1 & H_0 & \ddots  &  \\
&  \ddots & \ddots & H_1^{\dagger} \\
\lambda H_1^{\dagger}&  & H_1 & H_0
\end{pmatrix}
\notag \\
=&\sum_{x=-M}^{M}H_0\otimes \ket{x}\!\bra{x}+\sum_{x=-M}^{M-1}\Big{(}H_1\otimes \ket{x+1}\!\bra{x}+\textrm{H.c.}\Big{)} \notag \\
&\quad+\Big{(}\lambda H_1\otimes \ket{-M}\!\bra{M}+\textrm{H.c.} \Big{)}. \label{A1}
\end{align}
Here, $H_0$ and $H_1$ are $N_0\times N_0$ matrices, where $N_0$ is the number of states at each unit cell, coming from internal degrees of freedom. $H_0$ and $H_1$ represent the intra-unit-cell term and the nearest-neighbor hopping term, respectively. 
$\ket{x}$ is an eigenstate of the position operator $\hat{x}$ at the site $x$. 
Then $U_x\mathcal{H}(\lambda)U_x^{\dagger}$ is calculated as follows: 
\begin{align}
&\ U_x\mathcal{H}(\lambda)U_x^{\dagger} \notag \\
=&\sum_{x=-M}^{M}H_0\otimes \ket{x}\!\bra{x} \notag \\
&\quad+\sum_{x=-M}^{M-1}\Big{(}H_1\otimes \ket{x+1}\!\bra{x} e^{i\frac{\pi}{L}(x+1-x)}+ \textrm{H.c.} \Big{)} \notag \\
&\quad+\lambda \Big{(}H_1\otimes \ket{-M}\!\bra{M}e^{i\frac{\pi}{L}((-M)-M)}+\textrm{H.c.} \Big{)} \notag \\
=&\sum_{x=-M}^{M}H_0\otimes \ket{x}\!\bra{x} \notag \\
&\quad+\sum_{x=-M}^{M-1}\Big{(}(H_1e^{i\frac{\pi}{L}})\otimes \ket{x+1}\!\bra{x}+\textrm{H.c.}\Big{)} \notag \\
&\quad-\lambda \Big{(} (H_1e^{i\frac{\pi}{L}})\otimes \ket{-M}\!\bra{M} + \textrm{H.c.} \Big{)} \notag \\
=\ &\mathcal{H}(-\lambda)|_{H_1\to H_1e^{i\frac{\pi}{L}}}.
\end{align}
More generally, if we include a $m$th-nearest hopping term $\sum_{x}H_m\otimes\ket{x+m}\!\bra{x}+\textrm{H.c}$, we get 
\begin{align}
U_x\mathcal{H}(\lambda)U_x^{\dagger}
=\mathcal{H}(-\lambda)|_{H_m\to H_me^{i\frac{\pi m}{L}}}.
\label{A3}
\end{align}
Therefore, if $m$ is finite and $L$ is sufficiently large, $U_x\mathcal{H}(\lambda)U_x^{\dagger}$ is well-approximated by $\mathcal{H}(-\lambda)$.

\subsection{Anti-periodic Hamiltonian}

Here, we show that the Hamiltonian $\mathcal{H}|_{\lambda=-1}$ is unitary equivalent to the periodic Hamiltonian. 
For this purpose, one introduce the translational operator $T_x$ defined as $T_x=\sum_{x=-M}^{M-1}\ket{x+1}\!\bra{x}+\ket{-M}\!\bra{M}$. For $\lambda=1$, the Hamiltonian is expressed as follows: 
\begin{align}
&\quad\mathcal{H}|_{\lambda=1}\notag \\
&=H_0\otimes\sum_{x=-M}^{x=M}\ket{x}\!\bra{x}\notag \\
&\quad+H_1\otimes\Big{(}\sum_{x=-M}^{M-1}\ket{x+1}\!\bra{x}+\ket{-M}\!\bra{M}\Big{)}+\textrm{H.c.} \notag \\
&=H_0 \otimes \mathbf{1}_x + [H_1\otimes T_x + \textrm{H.c.}]. \label{A4}
\end{align}
For $\lambda=-1$, $U_x\mathcal{H}|_{\lambda=-1}U_x^{\dagger}$ is expressed as follows: 
\begin{align}
&\quad U_x\mathcal{H}|_{\lambda=-1}U_x^{\dagger} \notag \\
&=H_0\otimes\sum_{x=-M}^{x=M}\ket{x}\!\bra{x}\notag \\
&\quad+\Bigg{[}(H_1e^{i\frac{\pi}{L}})\otimes\Big{(}\sum_{x=-M}^{M-1}\ket{x+1}\!\bra{x}+\ket{-M}\!\bra{M}\Big{)}
+\textrm{H.c.}\Bigg{]} \notag \\
&=H_0 \otimes \mathbf{1}_x + [(H_1e^{i\frac{\pi}{L}})\otimes T_x + \textrm{H.c.}]
\notag \\
&=H_0 \otimes \mathbf{1}_x + [H_1\otimes \tilde{T_x} + \textrm{H.c.}], \label{A5}
\end{align}
where $\tilde{T_x}=e^{i\frac{\pi}{L}}T_x$. 
Let $e^{ik},e^{i\tilde{k}}$ be eigenvalues of $T_x,\tilde{T_x}$ respectively. 
$k,\tilde{k}$ take values as follows: 
\begin{align}
k&=\frac{2\pi}{L}m \quad(-M\leq m\leq M), \\
\tilde{k}&=\frac{2\pi}{L}m+\frac{\pi}{L} \quad(-M\leq m\leq M). 
\end{align}
By comparing Eqs.~(\ref{A4}) and (\ref{A5}), we conclude that $U_x\mathcal{H}|_{\lambda=-1}U_x^{\dagger}$ is equivalent to the Hamiltonian with the periodic boundary condition with $k$ shifted to $k+\pi/L$. 
This is natural from the following argument. The case with $\lambda=-1$ corresponds to antiperiodic boundary conditions, leading to the Aharonov–Bohm phase $\pi$ for the whole system. This additional phase $\pi$ appears as an additional term $\pi$ in the formula of $\tilde{k}L$. 

\subsection{Localized states and $U_x$}
Here, we show that if $\ket{\psi(\lambda)}$ is an eigenstate of $\mathcal{H}(\lambda)$ with parity $p\ (=\pm1)$ localized at the boundaries $x=\pm M$, then $U_x\ket{\psi(\lambda)}$ is an eigenstate of $\mathcal{H}(-\lambda)$ with parity $-p$ localized at the boundaries $x=\pm M$. 
That is because
\begin{align}
\mathcal{H}(-\lambda)U_x\ket{\psi(\lambda)}
&\simeq U_x\mathcal{H}(\lambda)U_x^{\dagger}U_x\ket{\psi(\lambda)} \notag \\
&=U_x\mathcal{H}\ket{\psi(\lambda)} \notag \\
&=E(\lambda)U_x\ket{\psi(\lambda)}, \label{A8}
\end{align}
\begin{align}
\mathcal{P}U_x\ket{\psi(\lambda)}
&=U_x^{\dagger}\mathcal{P}\ket{\psi(\lambda)} \notag \\
&=(U_x^{\dagger})^2 pU_x\ket{\psi(\lambda)} \notag \\
&\simeq(-p)U_x\ket{\psi(\lambda)}, \label{A9}
\end{align}
where $\simeq$ asymptotically holds true when $L\to\infty$. Here, we used the relation $(U_x^{\dagger})^2=\exp[-2i\pi x/L]\simeq-1\ (x\sim\pm M)$, which holds for localized states at the boundaries $x\simeq\pm M$. 
From Eqs.~(\ref{A8}) and (\ref{A9}), we conclude that if $\ket{\psi(\lambda_0)}$ is a boundary localized energy eigenstate with parity $p$, there is an eigenstate at $\lambda=-\lambda_0$, which have almost the same energy and have opposite parity $-p$. Therefore, if we label the localized states by an integer $l$, $\ket{\psi_l(\lambda)}$ and $\ket{\psi_l(-\lambda)}$ asymptotically have the same energy for a large system size, and have opposite parities.

\section{One-dimensional system with an even value of $L$}
In this section, we show that our theory is also applicable to a one-dimensional system with an even value of $L$. 
There are two different points from the case with an odd value of $L$. The first one is the inversion center for the inversion operation $\mathcal{P}$. 
Since the boundary point in the cutting procedure should be the inversion center, the other inversion center $x=0$ is at the center of a unit cell when $L$ is odd, and in between two unit cells when $L$ is even. Therefore, the definition of the bulk inversion operator $\mathcal{P}$ is different between odd $L$ and even $L$. Let $\mathcal{P}^{\text{odd}}$ ($\mathcal{P}^{\text{even}}$) denote the bulk inversion operators, with the inversion center being at the center of a unit cell (in the border between two neighboring unit cells). 
The following equations hold: 
\begin{align}
\mathcal{P}^{\text{even}}&=T_x \mathcal{P}^{\text{odd}}, \\
p^{\text{even}}(\Gamma_j)&=e^{-i \Gamma_j}p^{\text{odd}}(\Gamma_j),
\end{align}
where $p^{\text{even}}(\Gamma_j)$ ($p^{\text{odd}}(\Gamma_j)$) is an eigenvalue of $\mathcal{P}^{\text{even}}$ ($\mathcal{P}^{\text{odd}}$) at TRIM $\Gamma_j=0,\pi$. 
Therefore, 
\begin{align}
n_{\pm}^{\text{even}}(0)&=n_{\pm}^{\text{odd}}(0), \label{neno0} \\
n_{\pm}^{\text{even}}(\pi)&=n_{\mp}^{\text{odd}}(\pi), \notag \\
&=\nu-n_{\pm}^{\text{odd}}(\pi), \label{nenopi}
\end{align}
where $\nu$ is the number of bulk occupied bands.

\begin{figure}[t]
  \centerline{\includegraphics[width=6cm,clip]{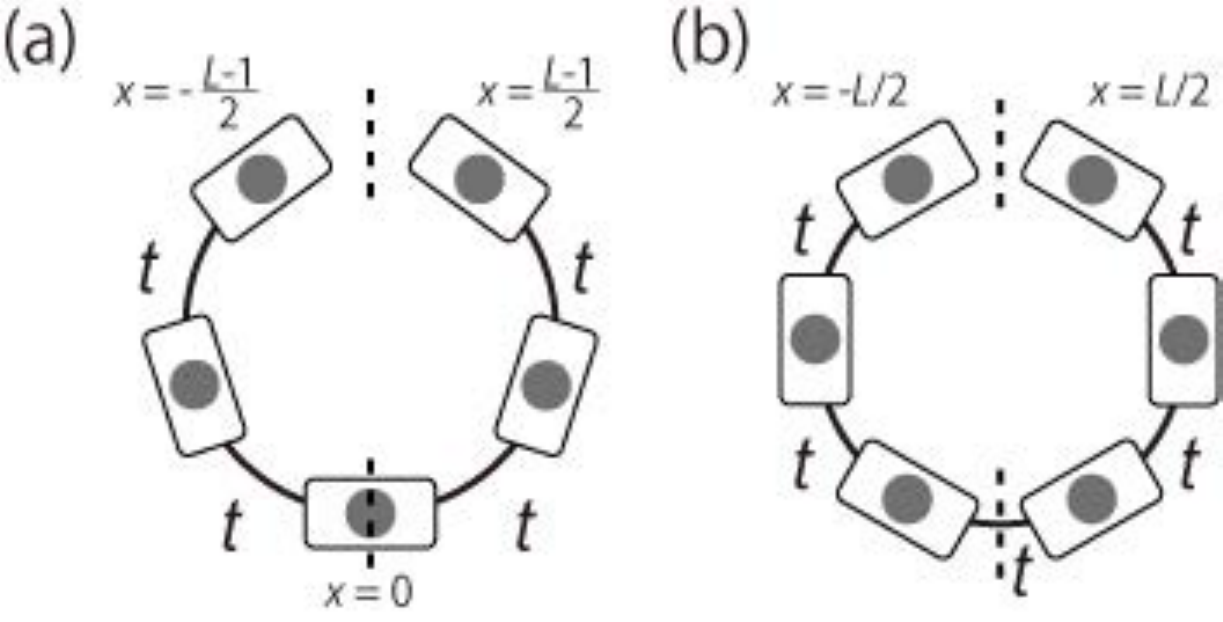}}
  \caption{Difference of the inversion center between (a) the case with odd $L$ and (b) the case with even $L$. 
}\label{even_odd}
\end{figure}

The other difference between cases with even $L$ and odd $L$ is the allowed values of $k$. For even $L$, $k=2\pi n/L$ can take both $0$ and $\pi$. On the other hand, $\tilde{k}=2\pi n/L+\pi/L$ does not take the values of the TRIM. Therefore, $N_{-}^{\text{even}}|_{\lambda=\pm1}$ are calculated as follows: 
\begin{align}
N_{-}^{\text{even}}|_{\lambda=1}&=\frac{(L-2)\nu}{2}+n_{-}^{\text{even}}(0)+n_{-}^{\text{even}}(\pi), \label{Neven0}
\\
N_{-}^{\text{even}}|_{\lambda=-1}&=\frac{L\nu}{2}. \label{Neven1}
\end{align}
Therefore, 
\begin{align}
N_{-}^{\text{even}}|_{\lambda=1} - N_{-}^{\text{even}}|_{\lambda=-1}
&=-\nu+n_{-}^{\text{even}}(0)+n_{-}^{\text{even}}(\pi). 
\end{align}
By using this result, the difference between $N|_{\lambda=1}$ and $N|_{\lambda=0}$ is calculated as follows:
\begin{align}
&N|_{\lambda=1}-N|_{\lambda=0}=[N]^{\lambda=1}_{\lambda=0}
\notag \\
=&[N_+^{\text{even}}]^{\lambda=1}_{\lambda=0}+[N_-^{\text{even}}]^{\lambda=1}_{\lambda=0}
\notag \\
=&[N_-^{\text{even}}]^{\lambda=-1}_{\lambda=0}+[N_-^{\text{even}}]^{\lambda=1}_{\lambda=0}
\notag \\
=&[N_-^{\text{even}}]^{\lambda=-1}_{\lambda=1}+2[N_-^{\text{even}}]^{\lambda=1}_{\lambda=0}
\notag \\
=&-\Big{(}-\nu+n_{-}^{\text{even}}(0)+n_{-}^{\text{even}}(\pi)\Big{)}
-2[N_-^{\text{even}}]^{\lambda=0}_{\lambda=1}, 
\end{align}
or equivalently, 
\begin{align}
&N|_{\lambda=0}
\notag \\
=&N|_{\lambda=1}
+\Big{(}-\nu+n_{-}^{\text{even}}(0)+n_{-}^{\text{even}}(\pi)\Big{)}
+2[N_-^{\text{even}}]^{\lambda=0}_{\lambda=1}. 
\label{NNeven}
\end{align}
By taking modulo 2 on both sides of the Eq.~(\ref{NNeven}), we get the following equation: 
\begin{align}
N|_{\lambda=0}\equiv N|_{\lambda=1}+
\Big{(}-\nu+n_{-}^{\text{even}}(0)+n_{-}^{\text{even}}(\pi)\Big{)}
\quad (\text{mod}\ 2).
\label{N0N1_even}
\end{align}
This means that the parity of the number of occupied states at $\lambda=0$ is calculated from the knowledge of $n^{\text{even}}_-$. 

Here, we show that this result gives the same value of $N|_{\lambda=0}$ as the case with odd $L$ in Eq.~(\ref{nuopen_nubulk}). 
From Eqs.~(\ref{neno0}) and (\ref{nenopi}), the second term in r.h.s. of Eq.~(\ref{N0N1_even}) is rewritten as follows: 
\begin{align}
&-\nu+n_{-}^{\text{even}}(0)+n_{-}^{\text{even}}(\pi) \notag \\
=&-\nu+n_{-}^{\text{odd}}(0)+n_{+}^{\text{odd}}(\pi) \notag \\
=&n_{-}^{\text{odd}}(0)-n_{-}^{\text{odd}}(\pi) 
\label{B10}.
\end{align}
From Eqs.~(\ref{N0N1_even}) and (\ref{B10}), we obtain Eq.~(\ref{nuopen_nubulk}). This clearly shows that the value of $N|_{\lambda=0}$ obtained here is the same as the case with odd $L$, which is physically reasonable.

\section{Three-dimensional system with even $L_1$}
In this section, we show that our theory is also applicable to a three-dimensional system with even $L_1$. 
The important point is that $\mu_1$ depends on the choice of the inversion center, if and only if $\nu_1\neq0$. Since the choice of the inversion center depends on the even-oddness of $L_1$ as explained in Appendix B, $\mu_1$ depends on $L_1$. Therefore, the Chern number of the system with open boundary also depends on $L_1$ when $\nu_1\neq0$. As shown below, the $L_1$ dependence of the Chern number is physically reasonable. 
This is the main focus of this section.

First, as in the case of odd $L_1$ studied in Sec.~I\hspace{-.1em}V A, we show that the Chern number (mod 2) of the system with open boundary condition is equal to the bulk topological number $\mu_1$. 
We write $n_-^{\text{even}}$ as $n_-^{\text{ev}}$ in this section. 
From Eqs.~(\ref{Neven0}) and (\ref{NNeven}), the following relation holds: 
\begin{align}
\tilde{n}^{\text{ev}}_-(\overline{\Gamma_j})|_{\lambda=1}&=\frac{(L_1-2)\nu}{2}+n^{\text{ev}}_{-}(0,\overline{\Gamma_j})+n^{\text{ev}}_{-}(\pi,\overline{\Gamma_j}), \label{NnG0_even}
\\
[\tilde{n}^{\text{ev}}_-(\overline{\Gamma_j})]^{\lambda=0}_{\lambda=1}
&=\frac{[N_{\overline{\Gamma_j}}]^{\lambda=0}_{\lambda=1}}{2}
+\frac{\nu-n^{\text{ev}}_-(0,\overline{\Gamma_j})-n^{\text{ev}}_-(\pi,\overline{\Gamma_j})}{2}. 
\label{nuopen_nubulk_g_even}
\end{align}
By adding Eq.~(\ref{NnG0_even}) and Eq.~(\ref{nuopen_nubulk_g_even}), we obtain the following equation: 
\begin{align}
\tilde{n}^{\text{ev}}_-(\overline{\Gamma_j})|_{\lambda=0}
&=\frac{1}{2}N_{\overline{\Gamma_j}}|_{\lambda=0}-\frac{\nu}{2}+\frac{n^{\text{ev}}_-(0,\overline{\Gamma_j})+n^{\text{ev}}_-(\pi,\overline{\Gamma_j})}{2}, \label{C10}
\end{align}
where we used the relation $N_{\overline{\Gamma_j}}|_{\lambda=0}=L\nu$. By taking the summation of Eq.~(\ref{C10}) over the 2D TRIM $\overline{\Gamma_j}$, we obtain the following relation: 
\begin{align}
&\sum_{\overline{\Gamma_j}}\tilde{n}^{\text{ev}}_-(\overline{\Gamma_j})|_{\lambda=0}
\notag \\
=&\frac{1}{2}\sum_{\overline{\Gamma_j}}N_{\overline{\Gamma_j}}|_{\lambda=0}
-2\nu+\sum_{\overline{\Gamma_j}}
\frac{n^{\text{ev}}_-(0,\overline{\Gamma_j})+n^{\text{ev}}_-(\pi,\overline{\Gamma_j})}{2}\label{C4}
\\
\equiv&\frac{1}{2}\sum_{\Gamma_j}n^{\text{ev}}_-(\Gamma_j) \quad (\text{mod}\ 2). 
\label{C5}
\end{align}
Equation (\ref{C5}) is completely the same form with Eq.~(\ref{Chern_mu_mod}), except for the definitions of $\tilde{n}_-$ and $n_-$: the parity is defined by $\mathcal{P}^{\text{odd}}$ in Eq.~(\ref{Chern_mu_mod}) and by $\mathcal{P}^{\text{even}}$ in (\ref{C5}). 
Here, $\mathcal{P}^{\text{odd}}$ and $\mathcal{P}^{\text{even}}$ are defined as in appendix B. 
Therefore, from Eqs.~(\ref{neno0}) and (\ref{nenopi}), Eqs.~(\ref{C4}) and (\ref{C5}) are rewritten in terms of as follows: 
\begin{align}
&\sum_{\overline{\Gamma_j}}\tilde{n}^{\text{ev}}_-(\overline{\Gamma_j})|_{\lambda=0}
\notag \\
=&\frac{1}{2}\sum_{\overline{\Gamma_j}}N_{\overline{\Gamma_j}}|_{\lambda=0}+\sum_{\overline{\Gamma_j}}
\frac{n^{\text{odd}}_-(0,\overline{\Gamma_j})-n^{\text{odd}}_-(\pi,\overline{\Gamma_j})}{2}
\\
\equiv&\frac{1}{2}\sum_{\Gamma_j}n^{\text{odd}}_-(\Gamma_j)-\nu_1
\quad (\text{mod}\ 2).
\label{C7} 
\end{align}
From Eqs.~(\ref{Chern_mu_mod}) and (\ref{C7}), we have the following relation:
\begin{align}
\sum_{\overline{\Gamma_j}}\tilde{n}^{\text{ev}}_-(\overline{\Gamma_j})|_{\lambda=0}
&\equiv
\sum_{\overline{\Gamma_j}}\tilde{n}^{\text{odd}}_-(\overline{\Gamma_j})|_{\lambda=0}
-\nu_1
\quad (\text{mod}\ 2), 
\\
(\text{Chern})^{\text{even}}|_{\lambda=0}
&\equiv
(\text{Chern})^{\text{odd}}|_{\lambda=0}
-\nu_1
\quad (\text{mod}\ 2). 
\label{C9}
\end{align}
Therefore, if $\nu_1\neq0$, the Chern number of the open 2D system depends on the even-oddness of $L_1$.

\begin{figure}[t]
  \centerline{\includegraphics[width=8cm,clip]{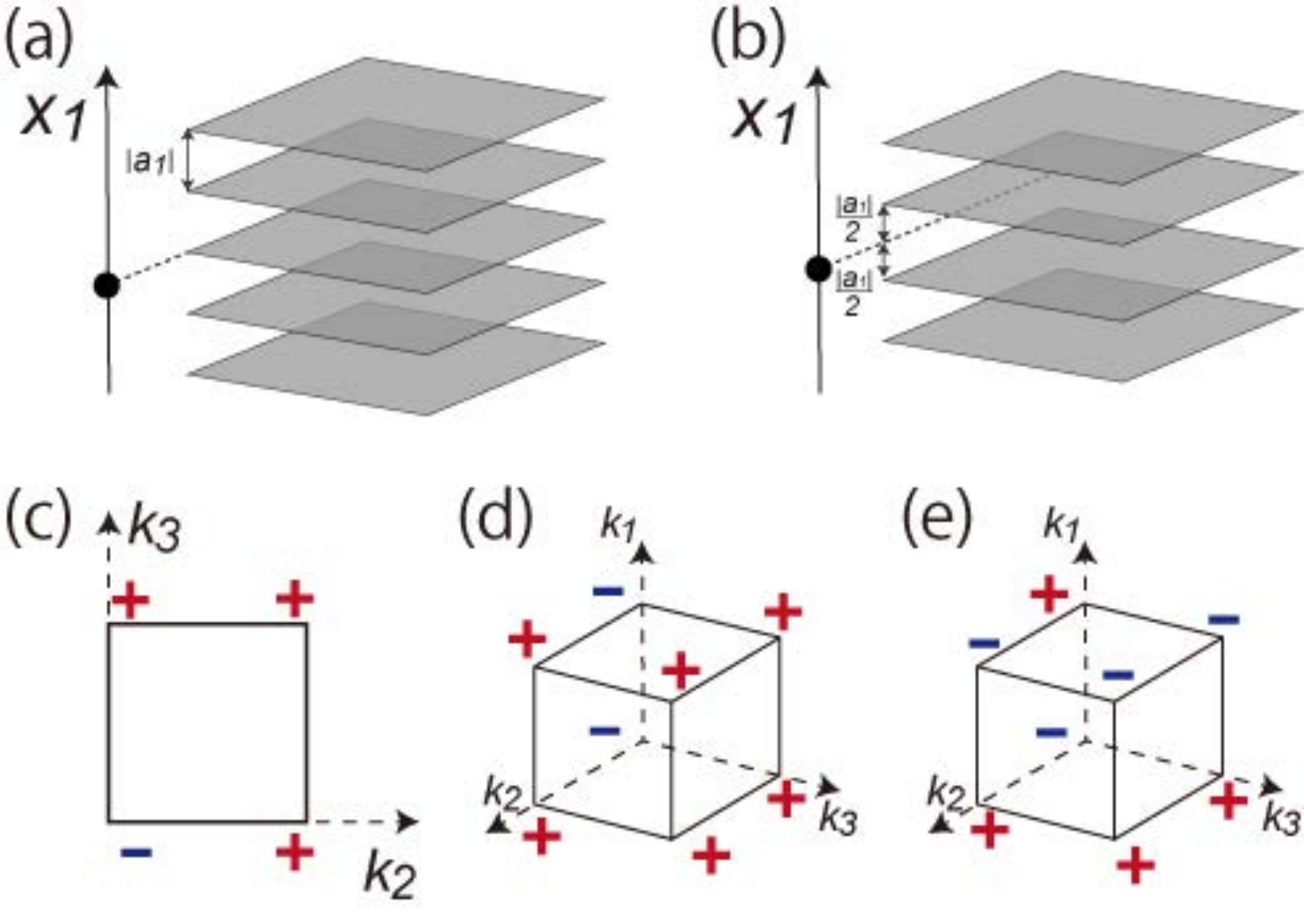}}
  \caption{(Color online) Conceptual figures of a stacking of $L_1$ layers of Chern insulators with (a) odd $L_1$ and (b) even $L_1$. Gray planes mean Chern insulator layers stacked along the $x_1$-direction. In (a), the inversion center is within one of the Chern insulator layer. In (b), the inversion center is in between the two layers. (c) An example of inversion parities of the Chern insulator layer. (d,e) The inversion parities of the stacked Chern insulators with (d) odd $L_1$ and (e) even $L_1$. The difference of the parities comes from the difference of the inversion center. 
}\label{weak_Chern_fig}
\end{figure}

Here, we give an illustrative example, in order to show that the $L_1$ dependence of the Chern number is physically reasonable.  
Let us consider a system formed by stacking 2D Chern insulators with Chern number $+1$. We assume that the stacking direction is the $x_1$ direction. In Fig.~\ref{weak_Chern_fig}, we illustrate the stacked Chern insulators with (a) an odd number of layers and (b) an even number of layers. 
For simplicity, we assume the inter-layer hopping term to be $0$. 
Then the 2D Chern number of the system perpendicular to $x_1$-axis is simply calculated as a sum of the Chern number of each layer, and therefore it is equal to the thickness of the layer $L_1$. In this case, obviously, the Chern number (mod 2) is equal to $L_1$ in accordance with Eq.~(\ref{C9})

Next, we additionally assume that the Chern insulator preserves inversion symmetry. Then, we can calculate the Chern number (mod 2) from the knowledge of the bulk inversion parity, via Eq.~(\ref{Chern_mu_mod}) or Eq.~(\ref{C5}). 
As an example, we consider the case that the parity eigenvalues of the Chern insulator are as shown in Fig.~\ref{weak_Chern_fig}(c), i.e. $n_-(\overline{\Gamma})=1$ and $n_-(\overline{\Gamma_j})=0$ ($\overline{\Gamma_j}=\overline{Y},\overline{Z},\overline{T}$). 
When we stack the Chern insulator with no inter-layer hopping, the bulk parity of the stacked 3D system at TRIM are directly calculated from the 2D bulk parity of the Chern insulator layer. The results depend on the even-oddness of the number of layers $L_1$, and shown in Fig.~\ref{weak_Chern_fig}(d,e). The $L_1$ dependence comes from the difference of the inversion center as shown in Fig.~\ref{weak_Chern_fig}(a,b). From Eq.~(\ref{Chern_mu_mod}) and Eq.~(\ref{C5}), the Chern number (mod 2) of the open system is calculated as $(\text{Chern})^{\text{odd}}|_{\lambda=0}\equiv1$ (mod 2) and $(\text{Chern})^{\text{even}}|_{\lambda=0}\equiv2\equiv0$ (mod 2), respectively. These results are consistent with the fact that the Chern number is equal to $L_1$.


\section{Boundary term}
In the main text, we only consider the simple open boundary condition. 
Here, we show that our results can also be applied to cases with other boundary conditions. 
We consider the boundary term defined as follows: 
\begin{align}
\mathcal{H}_{S}
&=
\begin{pmatrix}
\mathcal{H}^A & & & & \\
 & \bm{0} & & & \\
 & & \ddots  & &  \\
 & & & \bm{0} & \\
 & & & & \mathcal{H}^B
\end{pmatrix} 
\notag \\
&=\sum_{x,x'=0}^{l_A}\mathcal{H}^A_{x,x'} \ket{x}\!\bra{x'} \notag \\
&\quad\hphantom{=} +\sum_{x,x'=0}^{l_B}\mathcal{H}^B_{L-x,L-x'} \ket{L-x}\!\bra{L-x'},
\end{align}
where $l_A,l_B$ are small finite integers compared with the system size $L$. $\mathcal{H}^A_{x,x'}$ and $\mathcal{H}^B_{L-x,L-x'}$ are $N_0 \times N_0$ matrices which represent the change of the boundary condition. We assume that the boundary term does not break the inversion symmetry: $\mathcal{P}\mathcal{H}_S\mathcal{P}^{\dagger}=\mathcal{H}_S$. 
We add this term to the Hamiltonian with the simple boundary condition defined in Eq.~(\ref{A1}): $\mathcal{H}^{\text{basic}}(\lambda)\overset{\mathrm{def}}{=}\mathcal{H}(\lambda)$. Then the Hamiltonian with the general boundary condition $\mathcal{H}^{\text{general}}(\lambda)$ is defined as follows: 
\begin{align}
\mathcal{H}^{\text{general}}(\lambda)
&=\mathcal{H}^{\text{basic}}(\lambda)
+(1-\lambda^2)\mathcal{H}_{S}.
\end{align}
The additional term is added with a factor $1-\lambda^2$ in order to guarantee that it vanishes at $\lambda=\pm1$. In this section, we show that the additional term $(1-\lambda^2)\mathcal{H}_S$ does not change the results in the previous sections. 

First, the Hamiltonian $\mathcal{H}^{\text{general}}(\lambda)$ at $\lambda=\pm1$ are completely the same with $\mathcal{H}^{\text{basic}}(\lambda)$. Therefore, Eq.~(\ref{A5}) holds even when $\mathcal{H}(\lambda)$ is replaced by $\mathcal{H}^{\text{general}}(\lambda)$.
Next, we show that $U_x\mathcal{H}_SU_x^{\dagger}$ is well-approximated by $\mathcal{H}_S$:
\begin{align}
&\ U_x\mathcal{H}_SU_x^{\dagger}
\notag \\
&=\sum_{x,x'=0}^{l_A}\mathcal{H}^A_{x,x'} e^{\frac{i\pi}{L}(x-x')} \ket{x}\!\bra{x'} \notag \\
&\quad\hphantom{=} +\sum_{x,x'=0}^{l_B}\mathcal{H}^B_{L-x,L-x'}e^{-\frac{i\pi}{L}(x-x')} \ket{L-x}\!\bra{L-x'} 
\notag \\
&\simeq\mathcal{H}_S \quad(L\to\infty). 
\label{D3} 
\end{align}
Here, we used the fact that $e^{\frac{i\pi}{L}(x-x')}\to 1$ ($L\to\infty$), since $|x-x'|\leq l_A,l_B$ holds. 
From Eqs.~(\ref{A3}) and (\ref{D3}), we obtain the following relation:
\begin{align}
&\ U_x\mathcal{H}^{\text{general}}(\lambda)U_x^{\dagger}
\notag \\
&=U_x\mathcal{H}^{\text{basic}}(\lambda)U_x^{\dagger}
+(1-\lambda^2)U_x\mathcal{H}_SU_x^{\dagger}
\notag \\
&\simeq \mathcal{H}^{\text{basic}}(-\lambda)
+(1-(-\lambda)^2)\mathcal{H}_S
\notag \\
&=\mathcal{H}^{\text{general}}(-\lambda).
\label{D4}
\end{align}
By comparing Eq.~(\ref{A8}) and Eq.~(\ref{D4}), we can see that Eq.~(\ref{A8}) hold even when $\mathcal{H}(\lambda)$ is replaced by $\mathcal{H}^{\text{general}}(\lambda)$.

Combining the above two results, we conclude that our calculation holds true even when $\mathcal{H}(\lambda)$ is replaced by $\mathcal{H}^{\text{general}}(\lambda)$.


\begin{thebibliography}{52}%
\makeatletter
\providecommand \@ifxundefined [1]{%
 \@ifx{#1\undefined}
}%
\providecommand \@ifnum [1]{%
 \ifnum #1\expandafter \@firstoftwo
 \else \expandafter \@secondoftwo
 \fi
}%
\providecommand \@ifx [1]{%
 \ifx #1\expandafter \@firstoftwo
 \else \expandafter \@secondoftwo
 \fi
}%
\providecommand \natexlab [1]{#1}%
\providecommand \enquote  [1]{``#1''}%
\providecommand \bibnamefont  [1]{#1}%
\providecommand \bibfnamefont [1]{#1}%
\providecommand \citenamefont [1]{#1}%
\providecommand \href@noop [0]{\@secondoftwo}%
\providecommand \href [0]{\begingroup \@sanitize@url \@href}%
\providecommand \@href[1]{\@@startlink{#1}\@@href}%
\providecommand \@@href[1]{\endgroup#1\@@endlink}%
\providecommand \@sanitize@url [0]{\catcode `\\12\catcode `\$12\catcode
  `\&12\catcode `\#12\catcode `\^12\catcode `\_12\catcode `\%12\relax}%
\providecommand \@@startlink[1]{}%
\providecommand \@@endlink[0]{}%
\providecommand \url  [0]{\begingroup\@sanitize@url \@url }%
\providecommand \@url [1]{\endgroup\@href {#1}{\urlprefix }}%
\providecommand \urlprefix  [0]{URL }%
\providecommand \Eprint [0]{\href }%
\providecommand \doibase [0]{http://dx.doi.org/}%
\providecommand \selectlanguage [0]{\@gobble}%
\providecommand \bibinfo  [0]{\@secondoftwo}%
\providecommand \bibfield  [0]{\@secondoftwo}%
\providecommand \translation [1]{[#1]}%
\providecommand \BibitemOpen [0]{}%
\providecommand \bibitemStop [0]{}%
\providecommand \bibitemNoStop [0]{.\EOS\space}%
\providecommand \EOS [0]{\spacefactor3000\relax}%
\providecommand \BibitemShut  [1]{\csname bibitem#1\endcsname}%
\let\auto@bib@innerbib\@empty
\bibitem [{\citenamefont {Hasan}\ and\ \citenamefont
  {Kane}(2010)}]{RevModPhys.82.3045}%
  \BibitemOpen
  \bibfield  {author} {\bibinfo {author} {\bibfnamefont {M.~Z.}\ \bibnamefont
  {Hasan}}\ and\ \bibinfo {author} {\bibfnamefont {C.~L.}\ \bibnamefont
  {Kane}},\ }\bibfield  {title} {\enquote {\bibinfo {title}
  {\textit{Colloquium} : Topological insulators},}\ }\href
  {http://link.aps.org/doi/10.1103/RevModPhys.82.3045} {\bibfield  {journal}
  {\bibinfo  {journal} {Rev. Mod. Phys.}\ }\textbf {\bibinfo {volume} {82}},\
  \bibinfo {pages} {3045--3067} (\bibinfo {year} {2010})}\BibitemShut {NoStop}%
\bibitem [{\citenamefont {Qi}\ and\ \citenamefont
  {Zhang}(2011)}]{RevModPhys.83.1057}%
  \BibitemOpen
  \bibfield  {author} {\bibinfo {author} {\bibfnamefont {X.-L.}\ \bibnamefont
  {Qi}}\ and\ \bibinfo {author} {\bibfnamefont {S.-C.}\ \bibnamefont {Zhang}},\
  }\bibfield  {title} {\enquote {\bibinfo {title} {Topological insulators and
  superconductors},}\ }\href
  {http://link.aps.org/doi/10.1103/RevModPhys.83.1057} {\bibfield  {journal}
  {\bibinfo  {journal} {Rev. Mod. Phys.}\ }\textbf {\bibinfo {volume} {83}},\
  \bibinfo {pages} {1057--1110} (\bibinfo {year} {2011})}\BibitemShut {NoStop}%
\bibitem [{\citenamefont {Thouless}\ \emph {et~al.}(1982)\citenamefont
  {Thouless}, \citenamefont {Kohmoto}, \citenamefont {Nightingale},\ and\
  \citenamefont {den Nijs}}]{PhysRevLett.49.405}%
  \BibitemOpen
  \bibfield  {author} {\bibinfo {author} {\bibfnamefont {D.~J.}\ \bibnamefont
  {Thouless}}, \bibinfo {author} {\bibfnamefont {M.}~\bibnamefont {Kohmoto}},
  \bibinfo {author} {\bibfnamefont {M.~P.}\ \bibnamefont {Nightingale}}, \ and\
  \bibinfo {author} {\bibfnamefont {M.}~\bibnamefont {den Nijs}},\ }\bibfield
  {title} {\enquote {\bibinfo {title} {Quantized {Hall} conductance in a
  two-dimensional periodic potential},}\ }\href
  {https://link.aps.org/doi/10.1103/PhysRevLett.49.405} {\bibfield  {journal}
  {\bibinfo  {journal} {Phys. Rev. Lett.}\ }\textbf {\bibinfo {volume} {49}},\
  \bibinfo {pages} {405--408} (\bibinfo {year} {1982})}\BibitemShut {NoStop}%
\bibitem [{\citenamefont {Laughlin}(1981)}]{PhysRevB.23.5632}%
  \BibitemOpen
  \bibfield  {author} {\bibinfo {author} {\bibfnamefont {R.~B.}\ \bibnamefont
  {Laughlin}},\ }\bibfield  {title} {\enquote {\bibinfo {title} {Quantized
  {Hall} conductivity in two dimensions},}\ }\href
  {https://link.aps.org/doi/10.1103/PhysRevB.23.5632} {\bibfield  {journal}
  {\bibinfo  {journal} {Phys. Rev. B}\ }\textbf {\bibinfo {volume} {23}},\
  \bibinfo {pages} {5632--5633} (\bibinfo {year} {1981})}\BibitemShut {NoStop}%
\bibitem [{\citenamefont {Hatsugai}(1993)}]{PhysRevLett.71.3697}%
  \BibitemOpen
  \bibfield  {author} {\bibinfo {author} {\bibfnamefont {Y.}~\bibnamefont
  {Hatsugai}},\ }\bibfield  {title} {\enquote {\bibinfo {title} {Chern number
  and edge states in the integer quantum {Hall} effect},}\ }\href
  {https://link.aps.org/doi/10.1103/PhysRevLett.71.3697} {\bibfield  {journal}
  {\bibinfo  {journal} {Phys. Rev. Lett.}\ }\textbf {\bibinfo {volume} {71}},\
  \bibinfo {pages} {3697--3700} (\bibinfo {year} {1993})}\BibitemShut {NoStop}%
\bibitem [{\citenamefont {Kane}\ and\ \citenamefont
  {Mele}(2005{\natexlab{a}})}]{PhysRevLett.95.146802}%
  \BibitemOpen
  \bibfield  {author} {\bibinfo {author} {\bibfnamefont {C.~L.}\ \bibnamefont
  {Kane}}\ and\ \bibinfo {author} {\bibfnamefont {E.~J.}\ \bibnamefont
  {Mele}},\ }\bibfield  {title} {\enquote {\bibinfo {title} {${Z}_{2}$
  topological order and the quantum spin {Hall} effect},}\ }\href
  {http://link.aps.org/doi/10.1103/PhysRevLett.95.146802} {\bibfield  {journal}
  {\bibinfo  {journal} {Phys. Rev. Lett.}\ }\textbf {\bibinfo {volume} {95}},\
  \bibinfo {pages} {146802} (\bibinfo {year} {2005}{\natexlab{a}})}\BibitemShut
  {NoStop}%
\bibitem [{\citenamefont {Kane}\ and\ \citenamefont
  {Mele}(2005{\natexlab{b}})}]{PhysRevLett.95.226801}%
  \BibitemOpen
  \bibfield  {author} {\bibinfo {author} {\bibfnamefont {C.~L.}\ \bibnamefont
  {Kane}}\ and\ \bibinfo {author} {\bibfnamefont {E.~J.}\ \bibnamefont
  {Mele}},\ }\bibfield  {title} {\enquote {\bibinfo {title} {Quantum spin
  {Hall} effect in graphene},}\ }\href
  {http://link.aps.org/doi/10.1103/PhysRevLett.95.226801} {\bibfield  {journal}
  {\bibinfo  {journal} {Phys. Rev. Lett.}\ }\textbf {\bibinfo {volume} {95}},\
  \bibinfo {pages} {226801} (\bibinfo {year} {2005}{\natexlab{b}})}\BibitemShut
  {NoStop}%
\bibitem [{\citenamefont {Bernevig}\ and\ \citenamefont
  {Zhang}(2006)}]{PhysRevLett.96.106802}%
  \BibitemOpen
  \bibfield  {author} {\bibinfo {author} {\bibfnamefont {B.~A.}\ \bibnamefont
  {Bernevig}}\ and\ \bibinfo {author} {\bibfnamefont {S.-C.}\ \bibnamefont
  {Zhang}},\ }\bibfield  {title} {\enquote {\bibinfo {title} {Quantum spin
  {Hall} effect},}\ }\href
  {http://link.aps.org/doi/10.1103/PhysRevLett.96.106802} {\bibfield  {journal}
  {\bibinfo  {journal} {Phys. Rev. Lett.}\ }\textbf {\bibinfo {volume} {96}},\
  \bibinfo {pages} {106802} (\bibinfo {year} {2006})}\BibitemShut {NoStop}%
\bibitem [{\citenamefont {Zak}(1989)}]{PhysRevLett.62.2747}%
  \BibitemOpen
  \bibfield  {author} {\bibinfo {author} {\bibfnamefont {J.}~\bibnamefont
  {Zak}},\ }\bibfield  {title} {\enquote {\bibinfo {title} {Berry's phase for
  energy bands in solids},}\ }\href
  {https://link.aps.org/doi/10.1103/PhysRevLett.62.2747} {\bibfield  {journal}
  {\bibinfo  {journal} {Phys. Rev. Lett.}\ }\textbf {\bibinfo {volume} {62}},\
  \bibinfo {pages} {2747--2750} (\bibinfo {year} {1989})}\BibitemShut {NoStop}%
\bibitem [{\citenamefont {King-Smith}\ and\ \citenamefont
  {Vanderbilt}(1993)}]{PhysRevB.47.1651}%
  \BibitemOpen
  \bibfield  {author} {\bibinfo {author} {\bibfnamefont {R.~D.}\ \bibnamefont
  {King-Smith}}\ and\ \bibinfo {author} {\bibfnamefont {D.}~\bibnamefont
  {Vanderbilt}},\ }\bibfield  {title} {\enquote {\bibinfo {title} {Theory of
  polarization of crystalline solids},}\ }\href
  {https://link.aps.org/doi/10.1103/PhysRevB.47.1651} {\bibfield  {journal}
  {\bibinfo  {journal} {Phys. Rev. B}\ }\textbf {\bibinfo {volume} {47}},\
  \bibinfo {pages} {1651--1654} (\bibinfo {year} {1993})}\BibitemShut {NoStop}%
\bibitem [{\citenamefont {Vanderbilt}\ and\ \citenamefont
  {King-Smith}(1993)}]{PhysRevB.48.4442}%
  \BibitemOpen
  \bibfield  {author} {\bibinfo {author} {\bibfnamefont {David}\ \bibnamefont
  {Vanderbilt}}\ and\ \bibinfo {author} {\bibfnamefont {R.~D.}\ \bibnamefont
  {King-Smith}},\ }\bibfield  {title} {\enquote {\bibinfo {title} {Electric
  polarization as a bulk quantity and its relation to surface charge},}\ }\href
  {\doibase 10.1103/PhysRevB.48.4442} {\bibfield  {journal} {\bibinfo
  {journal} {Phys. Rev. B}\ }\textbf {\bibinfo {volume} {48}},\ \bibinfo
  {pages} {4442--4455} (\bibinfo {year} {1993})}\BibitemShut {NoStop}%
\bibitem [{\citenamefont {van Miert}\ \emph {et~al.}(2016)\citenamefont {van
  Miert}, \citenamefont {Ortix},\ and\ \citenamefont {Smith}}]{Miert_2016}%
  \BibitemOpen
  \bibfield  {author} {\bibinfo {author} {\bibfnamefont {Guido}\ \bibnamefont
  {van Miert}}, \bibinfo {author} {\bibfnamefont {Carmine}\ \bibnamefont
  {Ortix}}, \ and\ \bibinfo {author} {\bibfnamefont {Cristiane~Morais}\
  \bibnamefont {Smith}},\ }\bibfield  {title} {\enquote {\bibinfo {title}
  {Topological origin of edge states in two-dimensional inversion-symmetric
  insulators and semimetals},}\ }\href {\doibase 10.1088/2053-1583/4/1/015023}
  {\bibfield  {journal} {\bibinfo  {journal} {2D Materials}\ }\textbf {\bibinfo
  {volume} {4}},\ \bibinfo {pages} {015023} (\bibinfo {year}
  {2016})}\BibitemShut {NoStop}%
\bibitem [{\citenamefont {Hughes}\ \emph {et~al.}(2011)\citenamefont {Hughes},
  \citenamefont {Prodan},\ and\ \citenamefont {Bernevig}}]{PhysRevB.83.245132}%
  \BibitemOpen
  \bibfield  {author} {\bibinfo {author} {\bibfnamefont {Taylor~L.}\
  \bibnamefont {Hughes}}, \bibinfo {author} {\bibfnamefont {Emil}\ \bibnamefont
  {Prodan}}, \ and\ \bibinfo {author} {\bibfnamefont {B.~Andrei}\ \bibnamefont
  {Bernevig}},\ }\bibfield  {title} {\enquote {\bibinfo {title}
  {Inversion-symmetric topological insulators},}\ }\href {\doibase
  10.1103/PhysRevB.83.245132} {\bibfield  {journal} {\bibinfo  {journal} {Phys.
  Rev. B}\ }\textbf {\bibinfo {volume} {83}},\ \bibinfo {pages} {245132}
  (\bibinfo {year} {2011})}\BibitemShut {NoStop}%
\bibitem [{\citenamefont {Turner}\ \emph {et~al.}(2012)\citenamefont {Turner},
  \citenamefont {Zhang}, \citenamefont {Mong},\ and\ \citenamefont
  {Vishwanath}}]{PhysRevB.85.165120}%
  \BibitemOpen
  \bibfield  {author} {\bibinfo {author} {\bibfnamefont {Ari~M.}\ \bibnamefont
  {Turner}}, \bibinfo {author} {\bibfnamefont {Yi}~\bibnamefont {Zhang}},
  \bibinfo {author} {\bibfnamefont {Roger S.~K.}\ \bibnamefont {Mong}}, \ and\
  \bibinfo {author} {\bibfnamefont {Ashvin}\ \bibnamefont {Vishwanath}},\
  }\bibfield  {title} {\enquote {\bibinfo {title} {Quantized response and
  topology of magnetic insulators with inversion symmetry},}\ }\href {\doibase
  10.1103/PhysRevB.85.165120} {\bibfield  {journal} {\bibinfo  {journal} {Phys.
  Rev. B}\ }\textbf {\bibinfo {volume} {85}},\ \bibinfo {pages} {165120}
  (\bibinfo {year} {2012})}\BibitemShut {NoStop}%
\bibitem [{\citenamefont {Fang}\ \emph {et~al.}(2012)\citenamefont {Fang},
  \citenamefont {Gilbert},\ and\ \citenamefont
  {Bernevig}}]{PhysRevB.86.115112}%
  \BibitemOpen
  \bibfield  {author} {\bibinfo {author} {\bibfnamefont {C.}~\bibnamefont
  {Fang}}, \bibinfo {author} {\bibfnamefont {M.~J.}\ \bibnamefont {Gilbert}}, \
  and\ \bibinfo {author} {\bibfnamefont {B.~A.}\ \bibnamefont {Bernevig}},\
  }\bibfield  {title} {\enquote {\bibinfo {title} {Bulk topological invariants
  in noninteracting point group symmetric insulators},}\ }\href
  {https://link.aps.org/doi/10.1103/PhysRevB.86.115112} {\bibfield  {journal}
  {\bibinfo  {journal} {Phys. Rev. B}\ }\textbf {\bibinfo {volume} {86}},\
  \bibinfo {pages} {115112} (\bibinfo {year} {2012})}\BibitemShut {NoStop}%
\bibitem [{\citenamefont {Sitte}\ \emph {et~al.}(2012)\citenamefont {Sitte},
  \citenamefont {Rosch}, \citenamefont {Altman},\ and\ \citenamefont
  {Fritz}}]{PhysRevLett.108.126807}%
  \BibitemOpen
  \bibfield  {author} {\bibinfo {author} {\bibfnamefont {M.}~\bibnamefont
  {Sitte}}, \bibinfo {author} {\bibfnamefont {A.}~\bibnamefont {Rosch}},
  \bibinfo {author} {\bibfnamefont {E.}~\bibnamefont {Altman}}, \ and\ \bibinfo
  {author} {\bibfnamefont {L.}~\bibnamefont {Fritz}},\ }\bibfield  {title}
  {\enquote {\bibinfo {title} {Topological insulators in magnetic fields:
  Quantum hall effect and edge channels with a nonquantized
  $\ensuremath{\theta}$ term},}\ }\href {\doibase
  10.1103/PhysRevLett.108.126807} {\bibfield  {journal} {\bibinfo  {journal}
  {Phys. Rev. Lett.}\ }\textbf {\bibinfo {volume} {108}},\ \bibinfo {pages}
  {126807} (\bibinfo {year} {2012})}\BibitemShut {NoStop}%
\bibitem [{\citenamefont {Zhang}\ \emph {et~al.}(2013)\citenamefont {Zhang},
  \citenamefont {Kane},\ and\ \citenamefont {Mele}}]{PhysRevLett.110.046404}%
  \BibitemOpen
  \bibfield  {author} {\bibinfo {author} {\bibfnamefont {Fan}\ \bibnamefont
  {Zhang}}, \bibinfo {author} {\bibfnamefont {C.~L.}\ \bibnamefont {Kane}}, \
  and\ \bibinfo {author} {\bibfnamefont {E.~J.}\ \bibnamefont {Mele}},\
  }\bibfield  {title} {\enquote {\bibinfo {title} {Surface state magnetization
  and chiral edge states on topological insulators},}\ }\href {\doibase
  10.1103/PhysRevLett.110.046404} {\bibfield  {journal} {\bibinfo  {journal}
  {Phys. Rev. Lett.}\ }\textbf {\bibinfo {volume} {110}},\ \bibinfo {pages}
  {046404} (\bibinfo {year} {2013})}\BibitemShut {NoStop}%
\bibitem [{\citenamefont {Teo}\ and\ \citenamefont
  {Hughes}(2013)}]{PhysRevLett.111.047006}%
  \BibitemOpen
  \bibfield  {author} {\bibinfo {author} {\bibfnamefont {Jeffrey C.~Y.}\
  \bibnamefont {Teo}}\ and\ \bibinfo {author} {\bibfnamefont {Taylor~L.}\
  \bibnamefont {Hughes}},\ }\bibfield  {title} {\enquote {\bibinfo {title}
  {Existence of majorana-fermion bound states on disclinations and the
  classification of topological crystalline superconductors in two
  dimensions},}\ }\href {\doibase 10.1103/PhysRevLett.111.047006} {\bibfield
  {journal} {\bibinfo  {journal} {Phys. Rev. Lett.}\ }\textbf {\bibinfo
  {volume} {111}},\ \bibinfo {pages} {047006} (\bibinfo {year}
  {2013})}\BibitemShut {NoStop}%
\bibitem [{\citenamefont {Benalcazar}\ \emph {et~al.}(2014)\citenamefont
  {Benalcazar}, \citenamefont {Teo},\ and\ \citenamefont
  {Hughes}}]{PhysRevB.89.224503}%
  \BibitemOpen
  \bibfield  {author} {\bibinfo {author} {\bibfnamefont {Wladimir~A.}\
  \bibnamefont {Benalcazar}}, \bibinfo {author} {\bibfnamefont {Jeffrey C.~Y.}\
  \bibnamefont {Teo}}, \ and\ \bibinfo {author} {\bibfnamefont {Taylor~L.}\
  \bibnamefont {Hughes}},\ }\bibfield  {title} {\enquote {\bibinfo {title}
  {Classification of two-dimensional topological crystalline superconductors
  and majorana bound states at disclinations},}\ }\href {\doibase
  10.1103/PhysRevB.89.224503} {\bibfield  {journal} {\bibinfo  {journal} {Phys.
  Rev. B}\ }\textbf {\bibinfo {volume} {89}},\ \bibinfo {pages} {224503}
  (\bibinfo {year} {2014})}\BibitemShut {NoStop}%
\bibitem [{\citenamefont {Hashimoto}\ \emph {et~al.}(2017)\citenamefont
  {Hashimoto}, \citenamefont {Wu},\ and\ \citenamefont
  {Kimura}}]{PhysRevB.95.165443}%
  \BibitemOpen
  \bibfield  {author} {\bibinfo {author} {\bibfnamefont {Koji}\ \bibnamefont
  {Hashimoto}}, \bibinfo {author} {\bibfnamefont {Xi}~\bibnamefont {Wu}}, \
  and\ \bibinfo {author} {\bibfnamefont {Taro}\ \bibnamefont {Kimura}},\
  }\bibfield  {title} {\enquote {\bibinfo {title} {Edge states at an
  intersection of edges of a topological material},}\ }\href {\doibase
  10.1103/PhysRevB.95.165443} {\bibfield  {journal} {\bibinfo  {journal} {Phys.
  Rev. B}\ }\textbf {\bibinfo {volume} {95}},\ \bibinfo {pages} {165443}
  (\bibinfo {year} {2017})}\BibitemShut {NoStop}%
\bibitem [{\citenamefont {Benalcazar}\ \emph
  {et~al.}(2017{\natexlab{a}})\citenamefont {Benalcazar}, \citenamefont
  {Bernevig},\ and\ \citenamefont {Hughes}}]{Benalcazar61}%
  \BibitemOpen
  \bibfield  {author} {\bibinfo {author} {\bibfnamefont {Wladimir~A.}\
  \bibnamefont {Benalcazar}}, \bibinfo {author} {\bibfnamefont {B.~Andrei}\
  \bibnamefont {Bernevig}}, \ and\ \bibinfo {author} {\bibfnamefont
  {Taylor~L.}\ \bibnamefont {Hughes}},\ }\bibfield  {title} {\enquote {\bibinfo
  {title} {Quantized electric multipole insulators},}\ }\href {\doibase
  10.1126/science.aah6442} {\bibfield  {journal} {\bibinfo  {journal}
  {Science}\ }\textbf {\bibinfo {volume} {357}},\ \bibinfo {pages} {61--66}
  (\bibinfo {year} {2017}{\natexlab{a}})}\BibitemShut {NoStop}%
\bibitem [{\citenamefont {Fang}\ and\ \citenamefont {Fu}(2017)}]{fangc}%
  \BibitemOpen
  \bibfield  {author} {\bibinfo {author} {\bibfnamefont {C}~\bibnamefont
  {Fang}}\ and\ \bibinfo {author} {\bibfnamefont {L}~\bibnamefont {Fu}},\
  }\bibfield  {title} {\enquote {\bibinfo {title} {Rotation anomaly and
  topological crystalline insulators},}\ }\href
  {https://arxiv.org/abs/1709.01929} {\bibfield  {journal} {\bibinfo  {journal}
  {arXiv preprint arXiv:1709.01929}\ } (\bibinfo {year} {2017})}\BibitemShut
  {NoStop}%
\bibitem [{\citenamefont {Langbehn}\ \emph {et~al.}(2017)\citenamefont
  {Langbehn}, \citenamefont {Peng}, \citenamefont {Trifunovic}, \citenamefont
  {von Oppen},\ and\ \citenamefont {Brouwer}}]{PhysRevLett.119.246401}%
  \BibitemOpen
  \bibfield  {author} {\bibinfo {author} {\bibfnamefont {Josias}\ \bibnamefont
  {Langbehn}}, \bibinfo {author} {\bibfnamefont {Yang}\ \bibnamefont {Peng}},
  \bibinfo {author} {\bibfnamefont {Luka}\ \bibnamefont {Trifunovic}}, \bibinfo
  {author} {\bibfnamefont {Felix}\ \bibnamefont {von Oppen}}, \ and\ \bibinfo
  {author} {\bibfnamefont {Piet~W.}\ \bibnamefont {Brouwer}},\ }\bibfield
  {title} {\enquote {\bibinfo {title} {Reflection-symmetric second-order
  topological insulators and superconductors},}\ }\href {\doibase
  10.1103/PhysRevLett.119.246401} {\bibfield  {journal} {\bibinfo  {journal}
  {Phys. Rev. Lett.}\ }\textbf {\bibinfo {volume} {119}},\ \bibinfo {pages}
  {246401} (\bibinfo {year} {2017})}\BibitemShut {NoStop}%
\bibitem [{\citenamefont {Song}\ \emph {et~al.}(2017)\citenamefont {Song},
  \citenamefont {Fang},\ and\ \citenamefont {Fang}}]{PhysRevLett.119.246402}%
  \BibitemOpen
  \bibfield  {author} {\bibinfo {author} {\bibfnamefont {Zhida}\ \bibnamefont
  {Song}}, \bibinfo {author} {\bibfnamefont {Zhong}\ \bibnamefont {Fang}}, \
  and\ \bibinfo {author} {\bibfnamefont {Chen}\ \bibnamefont {Fang}},\
  }\bibfield  {title} {\enquote {\bibinfo {title}
  {$(d\ensuremath{-}2)$-dimensional edge states of rotation symmetry protected
  topological states},}\ }\href {\doibase 10.1103/PhysRevLett.119.246402}
  {\bibfield  {journal} {\bibinfo  {journal} {Phys. Rev. Lett.}\ }\textbf
  {\bibinfo {volume} {119}},\ \bibinfo {pages} {246402} (\bibinfo {year}
  {2017})}\BibitemShut {NoStop}%
\bibitem [{\citenamefont {Benalcazar}\ \emph
  {et~al.}(2017{\natexlab{b}})\citenamefont {Benalcazar}, \citenamefont
  {Bernevig},\ and\ \citenamefont {Hughes}}]{PhysRevB.96.245115}%
  \BibitemOpen
  \bibfield  {author} {\bibinfo {author} {\bibfnamefont {Wladimir~A.}\
  \bibnamefont {Benalcazar}}, \bibinfo {author} {\bibfnamefont {B.~Andrei}\
  \bibnamefont {Bernevig}}, \ and\ \bibinfo {author} {\bibfnamefont
  {Taylor~L.}\ \bibnamefont {Hughes}},\ }\bibfield  {title} {\enquote {\bibinfo
  {title} {Electric multipole moments, topological multipole moment pumping,
  and chiral hinge states in crystalline insulators},}\ }\href {\doibase
  10.1103/PhysRevB.96.245115} {\bibfield  {journal} {\bibinfo  {journal} {Phys.
  Rev. B}\ }\textbf {\bibinfo {volume} {96}},\ \bibinfo {pages} {245115}
  (\bibinfo {year} {2017}{\natexlab{b}})}\BibitemShut {NoStop}%
\bibitem [{\citenamefont {Serra-Garcia}\ \emph {et~al.}(2018)\citenamefont
  {Serra-Garcia}, \citenamefont {Peri}, \citenamefont {S{\"u}sstrunk},
  \citenamefont {Bilal}, \citenamefont {Larsen}, \citenamefont {Villanueva},\
  and\ \citenamefont {Huber}}]{serra2018observation}%
  \BibitemOpen
  \bibfield  {author} {\bibinfo {author} {\bibfnamefont {Marc}\ \bibnamefont
  {Serra-Garcia}}, \bibinfo {author} {\bibfnamefont {Valerio}\ \bibnamefont
  {Peri}}, \bibinfo {author} {\bibfnamefont {Roman}\ \bibnamefont
  {S{\"u}sstrunk}}, \bibinfo {author} {\bibfnamefont {Osama~R}\ \bibnamefont
  {Bilal}}, \bibinfo {author} {\bibfnamefont {Tom}\ \bibnamefont {Larsen}},
  \bibinfo {author} {\bibfnamefont {Luis~Guillermo}\ \bibnamefont
  {Villanueva}}, \ and\ \bibinfo {author} {\bibfnamefont {Sebastian~D}\
  \bibnamefont {Huber}},\ }\bibfield  {title} {\enquote {\bibinfo {title}
  {Observation of a phononic quadrupole topological insulator},}\ }\href
  {https://doi.org/10.1038/nature25156} {\bibfield  {journal} {\bibinfo
  {journal} {Nature}\ }\textbf {\bibinfo {volume} {555}},\ \bibinfo {pages}
  {342} (\bibinfo {year} {2018})}\BibitemShut {NoStop}%
\bibitem [{\citenamefont {Ezawa}(2018{\natexlab{a}})}]{PhysRevLett.120.026801}%
  \BibitemOpen
  \bibfield  {author} {\bibinfo {author} {\bibfnamefont {Motohiko}\
  \bibnamefont {Ezawa}},\ }\bibfield  {title} {\enquote {\bibinfo {title}
  {Higher-order topological insulators and semimetals on the breathing kagome
  and pyrochlore lattices},}\ }\href {\doibase 10.1103/PhysRevLett.120.026801}
  {\bibfield  {journal} {\bibinfo  {journal} {Phys. Rev. Lett.}\ }\textbf
  {\bibinfo {volume} {120}},\ \bibinfo {pages} {026801} (\bibinfo {year}
  {2018}{\natexlab{a}})}\BibitemShut {NoStop}%
\bibitem [{\citenamefont {Peterson}\ \emph {et~al.}(2018)\citenamefont
  {Peterson}, \citenamefont {Benalcazar}, \citenamefont {Hughes},\ and\
  \citenamefont {Bahl}}]{peterson2018quantized}%
  \BibitemOpen
  \bibfield  {author} {\bibinfo {author} {\bibfnamefont {Christopher~W}\
  \bibnamefont {Peterson}}, \bibinfo {author} {\bibfnamefont {Wladimir~A}\
  \bibnamefont {Benalcazar}}, \bibinfo {author} {\bibfnamefont {Taylor~L}\
  \bibnamefont {Hughes}}, \ and\ \bibinfo {author} {\bibfnamefont {Gaurav}\
  \bibnamefont {Bahl}},\ }\bibfield  {title} {\enquote {\bibinfo {title} {A
  quantized microwave quadrupole insulator with topologically protected corner
  states},}\ }\href {https://doi.org/10.1038/nature25777} {\bibfield  {journal}
  {\bibinfo  {journal} {Nature}\ }\textbf {\bibinfo {volume} {555}},\ \bibinfo
  {pages} {346} (\bibinfo {year} {2018})}\BibitemShut {NoStop}%
\bibitem [{\citenamefont {Ezawa}(2018{\natexlab{b}})}]{PhysRevB.97.155305}%
  \BibitemOpen
  \bibfield  {author} {\bibinfo {author} {\bibfnamefont {Motohiko}\
  \bibnamefont {Ezawa}},\ }\bibfield  {title} {\enquote {\bibinfo {title}
  {Magnetic second-order topological insulators and semimetals},}\ }\href
  {\doibase 10.1103/PhysRevB.97.155305} {\bibfield  {journal} {\bibinfo
  {journal} {Phys. Rev. B}\ }\textbf {\bibinfo {volume} {97}},\ \bibinfo
  {pages} {155305} (\bibinfo {year} {2018}{\natexlab{b}})}\BibitemShut
  {NoStop}%
\bibitem [{\citenamefont {Ezawa}(2018{\natexlab{c}})}]{PhysRevB.97.241402}%
  \BibitemOpen
  \bibfield  {author} {\bibinfo {author} {\bibfnamefont {Motohiko}\
  \bibnamefont {Ezawa}},\ }\bibfield  {title} {\enquote {\bibinfo {title}
  {Strong and weak second-order topological insulators with hexagonal symmetry
  and \ensuremath{\mathbb{Z}}${}_{3}$ index},}\ }\href {\doibase
  10.1103/PhysRevB.97.241402} {\bibfield  {journal} {\bibinfo  {journal} {Phys.
  Rev. B}\ }\textbf {\bibinfo {volume} {97}},\ \bibinfo {pages} {241402}
  (\bibinfo {year} {2018}{\natexlab{c}})}\BibitemShut {NoStop}%
\bibitem [{\citenamefont {Geier}\ \emph {et~al.}(2018)\citenamefont {Geier},
  \citenamefont {Trifunovic}, \citenamefont {Hoskam},\ and\ \citenamefont
  {Brouwer}}]{PhysRevB.97.205135}%
  \BibitemOpen
  \bibfield  {author} {\bibinfo {author} {\bibfnamefont {Max}\ \bibnamefont
  {Geier}}, \bibinfo {author} {\bibfnamefont {Luka}\ \bibnamefont
  {Trifunovic}}, \bibinfo {author} {\bibfnamefont {Max}\ \bibnamefont
  {Hoskam}}, \ and\ \bibinfo {author} {\bibfnamefont {Piet~W.}\ \bibnamefont
  {Brouwer}},\ }\bibfield  {title} {\enquote {\bibinfo {title} {Second-order
  topological insulators and superconductors with an order-two crystalline
  symmetry},}\ }\href {\doibase 10.1103/PhysRevB.97.205135} {\bibfield
  {journal} {\bibinfo  {journal} {Phys. Rev. B}\ }\textbf {\bibinfo {volume}
  {97}},\ \bibinfo {pages} {205135} (\bibinfo {year} {2018})}\BibitemShut
  {NoStop}%
\bibitem [{\citenamefont {Khalaf}(2018)}]{PhysRevB.97.205136}%
  \BibitemOpen
  \bibfield  {author} {\bibinfo {author} {\bibfnamefont {Eslam}\ \bibnamefont
  {Khalaf}},\ }\bibfield  {title} {\enquote {\bibinfo {title} {Higher-order
  topological insulators and superconductors protected by inversion
  symmetry},}\ }\href {\doibase 10.1103/PhysRevB.97.205136} {\bibfield
  {journal} {\bibinfo  {journal} {Phys. Rev. B}\ }\textbf {\bibinfo {volume}
  {97}},\ \bibinfo {pages} {205136} (\bibinfo {year} {2018})}\BibitemShut
  {NoStop}%
\bibitem [{\citenamefont {Schindler}\ \emph
  {et~al.}(2018{\natexlab{a}})\citenamefont {Schindler}, \citenamefont {Cook},
  \citenamefont {Vergniory}, \citenamefont {Wang}, \citenamefont {Parkin},
  \citenamefont {Bernevig},\ and\ \citenamefont
  {Neupert}}]{schindler2018higherTI}%
  \BibitemOpen
  \bibfield  {author} {\bibinfo {author} {\bibfnamefont {Frank}\ \bibnamefont
  {Schindler}}, \bibinfo {author} {\bibfnamefont {Ashley~M}\ \bibnamefont
  {Cook}}, \bibinfo {author} {\bibfnamefont {Maia~G}\ \bibnamefont
  {Vergniory}}, \bibinfo {author} {\bibfnamefont {Zhijun}\ \bibnamefont
  {Wang}}, \bibinfo {author} {\bibfnamefont {Stuart~SP}\ \bibnamefont
  {Parkin}}, \bibinfo {author} {\bibfnamefont {B~Andrei}\ \bibnamefont
  {Bernevig}}, \ and\ \bibinfo {author} {\bibfnamefont {Titus}\ \bibnamefont
  {Neupert}},\ }\bibfield  {title} {\enquote {\bibinfo {title} {Higher-order
  topological insulators},}\ }\href
  {http://advances.sciencemag.org/content/4/6/eaat0346} {\bibfield  {journal}
  {\bibinfo  {journal} {Sci. Adv.}\ }\textbf {\bibinfo {volume} {4}},\ \bibinfo
  {pages} {eaat0346} (\bibinfo {year} {2018}{\natexlab{a}})}\BibitemShut
  {NoStop}%
\bibitem [{\citenamefont {Ezawa}(2018{\natexlab{d}})}]{PhysRevB.98.045125}%
  \BibitemOpen
  \bibfield  {author} {\bibinfo {author} {\bibfnamefont {Motohiko}\
  \bibnamefont {Ezawa}},\ }\bibfield  {title} {\enquote {\bibinfo {title}
  {Minimal models for wannier-type higher-order topological insulators and
  phosphorene},}\ }\href {\doibase 10.1103/PhysRevB.98.045125} {\bibfield
  {journal} {\bibinfo  {journal} {Phys. Rev. B}\ }\textbf {\bibinfo {volume}
  {98}},\ \bibinfo {pages} {045125} (\bibinfo {year}
  {2018}{\natexlab{d}})}\BibitemShut {NoStop}%
\bibitem [{\citenamefont {Schindler}\ \emph
  {et~al.}(2018{\natexlab{b}})\citenamefont {Schindler}, \citenamefont {Wang},
  \citenamefont {Vergniory}, \citenamefont {Cook}, \citenamefont {Murani},
  \citenamefont {Sengupta}, \citenamefont {Kasumov}, \citenamefont {Deblock},
  \citenamefont {Jeon}, \citenamefont {Drozdov} \emph
  {et~al.}}]{schindler2018higher}%
  \BibitemOpen
  \bibfield  {author} {\bibinfo {author} {\bibfnamefont {Frank}\ \bibnamefont
  {Schindler}}, \bibinfo {author} {\bibfnamefont {Zhijun}\ \bibnamefont
  {Wang}}, \bibinfo {author} {\bibfnamefont {Maia~G}\ \bibnamefont
  {Vergniory}}, \bibinfo {author} {\bibfnamefont {Ashley~M}\ \bibnamefont
  {Cook}}, \bibinfo {author} {\bibfnamefont {Anil}\ \bibnamefont {Murani}},
  \bibinfo {author} {\bibfnamefont {Shamashis}\ \bibnamefont {Sengupta}},
  \bibinfo {author} {\bibfnamefont {Alik~Yu}\ \bibnamefont {Kasumov}}, \bibinfo
  {author} {\bibfnamefont {Richard}\ \bibnamefont {Deblock}}, \bibinfo {author}
  {\bibfnamefont {Sangjun}\ \bibnamefont {Jeon}}, \bibinfo {author}
  {\bibfnamefont {Ilya}\ \bibnamefont {Drozdov}},  \emph {et~al.},\ }\bibfield
  {title} {\enquote {\bibinfo {title} {Higher-order topology in bismuth},}\
  }\href {https://doi.org/10.1038/s41567-018-0224-7} {\bibfield  {journal}
  {\bibinfo  {journal} {Nature Physics}\ }\textbf {\bibinfo {volume} {14}},\
  \bibinfo {pages} {918} (\bibinfo {year} {2018}{\natexlab{b}})}\BibitemShut
  {NoStop}%
\bibitem [{\citenamefont {van Miert}\ and\ \citenamefont
  {Ortix}(2018)}]{PhysRevB.98.081110}%
  \BibitemOpen
  \bibfield  {author} {\bibinfo {author} {\bibfnamefont {Guido}\ \bibnamefont
  {van Miert}}\ and\ \bibinfo {author} {\bibfnamefont {Carmine}\ \bibnamefont
  {Ortix}},\ }\bibfield  {title} {\enquote {\bibinfo {title} {Higher-order
  topological insulators protected by inversion and rotoinversion
  symmetries},}\ }\href {\doibase 10.1103/PhysRevB.98.081110} {\bibfield
  {journal} {\bibinfo  {journal} {Phys. Rev. B}\ }\textbf {\bibinfo {volume}
  {98}},\ \bibinfo {pages} {081110} (\bibinfo {year} {2018})}\BibitemShut
  {NoStop}%
\bibitem [{\citenamefont {Khalaf}\ \emph {et~al.}(2018)\citenamefont {Khalaf},
  \citenamefont {Po}, \citenamefont {Vishwanath},\ and\ \citenamefont
  {Watanabe}}]{PhysRevX.8.031070}%
  \BibitemOpen
  \bibfield  {author} {\bibinfo {author} {\bibfnamefont {Eslam}\ \bibnamefont
  {Khalaf}}, \bibinfo {author} {\bibfnamefont {Hoi~Chun}\ \bibnamefont {Po}},
  \bibinfo {author} {\bibfnamefont {Ashvin}\ \bibnamefont {Vishwanath}}, \ and\
  \bibinfo {author} {\bibfnamefont {Haruki}\ \bibnamefont {Watanabe}},\
  }\bibfield  {title} {\enquote {\bibinfo {title} {Symmetry indicators and
  anomalous surface states of topological crystalline insulators},}\ }\href
  {\doibase 10.1103/PhysRevX.8.031070} {\bibfield  {journal} {\bibinfo
  {journal} {Phys. Rev. X}\ }\textbf {\bibinfo {volume} {8}},\ \bibinfo {pages}
  {031070} (\bibinfo {year} {2018})}\BibitemShut {NoStop}%
\bibitem [{\citenamefont {Imhof}\ \emph {et~al.}(2018)\citenamefont {Imhof},
  \citenamefont {Berger}, \citenamefont {Bayer}, \citenamefont {Brehm},
  \citenamefont {Molenkamp}, \citenamefont {Kiessling}, \citenamefont
  {Schindler}, \citenamefont {Lee}, \citenamefont {Greiter}, \citenamefont
  {Neupert} \emph {et~al.}}]{imhof2018topolectrical}%
  \BibitemOpen
  \bibfield  {author} {\bibinfo {author} {\bibfnamefont {Stefan}\ \bibnamefont
  {Imhof}}, \bibinfo {author} {\bibfnamefont {Christian}\ \bibnamefont
  {Berger}}, \bibinfo {author} {\bibfnamefont {Florian}\ \bibnamefont {Bayer}},
  \bibinfo {author} {\bibfnamefont {Johannes}\ \bibnamefont {Brehm}}, \bibinfo
  {author} {\bibfnamefont {Laurens~W}\ \bibnamefont {Molenkamp}}, \bibinfo
  {author} {\bibfnamefont {Tobias}\ \bibnamefont {Kiessling}}, \bibinfo
  {author} {\bibfnamefont {Frank}\ \bibnamefont {Schindler}}, \bibinfo {author}
  {\bibfnamefont {Ching~Hua}\ \bibnamefont {Lee}}, \bibinfo {author}
  {\bibfnamefont {Martin}\ \bibnamefont {Greiter}}, \bibinfo {author}
  {\bibfnamefont {Titus}\ \bibnamefont {Neupert}},  \emph {et~al.},\ }\bibfield
   {title} {\enquote {\bibinfo {title} {Topolectrical-circuit realization of
  topological corner modes},}\ }\href
  {https://doi.org/10.1038/s41567-018-0246-1} {\bibfield  {journal} {\bibinfo
  {journal} {Nature Physics}\ }\textbf {\bibinfo {volume} {14}},\ \bibinfo
  {pages} {925} (\bibinfo {year} {2018})}\BibitemShut {NoStop}%
\bibitem [{\citenamefont {Matsugatani}\ and\ \citenamefont
  {Watanabe}(2018)}]{PhysRevB.98.205129}%
  \BibitemOpen
  \bibfield  {author} {\bibinfo {author} {\bibfnamefont {Akishi}\ \bibnamefont
  {Matsugatani}}\ and\ \bibinfo {author} {\bibfnamefont {Haruki}\ \bibnamefont
  {Watanabe}},\ }\bibfield  {title} {\enquote {\bibinfo {title} {Connecting
  higher-order topological insulators to lower-dimensional topological
  insulators},}\ }\href {\doibase 10.1103/PhysRevB.98.205129} {\bibfield
  {journal} {\bibinfo  {journal} {Phys. Rev. B}\ }\textbf {\bibinfo {volume}
  {98}},\ \bibinfo {pages} {205129} (\bibinfo {year} {2018})}\BibitemShut
  {NoStop}%
\bibitem [{\citenamefont {Kooi}\ \emph {et~al.}(2018)\citenamefont {Kooi},
  \citenamefont {van Miert},\ and\ \citenamefont {Ortix}}]{PhysRevB.98.245102}%
  \BibitemOpen
  \bibfield  {author} {\bibinfo {author} {\bibfnamefont {Sander~H.}\
  \bibnamefont {Kooi}}, \bibinfo {author} {\bibfnamefont {Guido}\ \bibnamefont
  {van Miert}}, \ and\ \bibinfo {author} {\bibfnamefont {Carmine}\ \bibnamefont
  {Ortix}},\ }\bibfield  {title} {\enquote {\bibinfo {title}
  {Inversion-symmetry protected chiral hinge states in stacks of doped quantum
  hall layers},}\ }\href {\doibase 10.1103/PhysRevB.98.245102} {\bibfield
  {journal} {\bibinfo  {journal} {Phys. Rev. B}\ }\textbf {\bibinfo {volume}
  {98}},\ \bibinfo {pages} {245102} (\bibinfo {year} {2018})}\BibitemShut
  {NoStop}%
\bibitem [{\citenamefont {Trifunovic}\ and\ \citenamefont
  {Brouwer}(2019)}]{PhysRevX.9.011012}%
  \BibitemOpen
  \bibfield  {author} {\bibinfo {author} {\bibfnamefont {Luka}\ \bibnamefont
  {Trifunovic}}\ and\ \bibinfo {author} {\bibfnamefont {Piet~W.}\ \bibnamefont
  {Brouwer}},\ }\bibfield  {title} {\enquote {\bibinfo {title} {Higher-order
  bulk-boundary correspondence for topological crystalline phases},}\ }\href
  {\doibase 10.1103/PhysRevX.9.011012} {\bibfield  {journal} {\bibinfo
  {journal} {Phys. Rev. X}\ }\textbf {\bibinfo {volume} {9}},\ \bibinfo {pages}
  {011012} (\bibinfo {year} {2019})}\BibitemShut {NoStop}%
\bibitem [{\citenamefont {Teo}\ \emph {et~al.}(2008)\citenamefont {Teo},
  \citenamefont {Fu},\ and\ \citenamefont {Kane}}]{PhysRevB.78.045426}%
  \BibitemOpen
  \bibfield  {author} {\bibinfo {author} {\bibfnamefont {Jeffrey C.~Y.}\
  \bibnamefont {Teo}}, \bibinfo {author} {\bibfnamefont {Liang}\ \bibnamefont
  {Fu}}, \ and\ \bibinfo {author} {\bibfnamefont {C.~L.}\ \bibnamefont
  {Kane}},\ }\bibfield  {title} {\enquote {\bibinfo {title} {Surface states and
  topological invariants in three-dimensional topological insulators:
  Application to ${\text{bi}}_{1\ensuremath{-}x}{\text{sb}}_{x}$},}\ }\href
  {\doibase 10.1103/PhysRevB.78.045426} {\bibfield  {journal} {\bibinfo
  {journal} {Phys. Rev. B}\ }\textbf {\bibinfo {volume} {78}},\ \bibinfo
  {pages} {045426} (\bibinfo {year} {2008})}\BibitemShut {NoStop}%
\bibitem [{\citenamefont {Tanaka}\ \emph {et~al.}(2019)\citenamefont {Tanaka},
  \citenamefont {Takahashi},\ and\ \citenamefont
  {Murakami}}]{Tanaka_preparation}%
  \BibitemOpen
  \bibfield  {author} {\bibinfo {author} {\bibfnamefont {Yutaro}\ \bibnamefont
  {Tanaka}}, \bibinfo {author} {\bibfnamefont {Ryo}\ \bibnamefont {Takahashi}},
  \ and\ \bibinfo {author} {\bibfnamefont {Shuichi}\ \bibnamefont {Murakami}},\
  }\bibfield  {title} {\enquote {\bibinfo {title} {Appearance of hinge states
  in second-order topological insulators via the cutting procedure},}\ }\href
  {https://arxiv.org/abs/1910.05938} {\bibfield  {journal} {\bibinfo  {journal}
  {arXiv preprint arXiv:1910.05938}\ } (\bibinfo {year} {2019})}\BibitemShut
  {NoStop}%
\bibitem [{\citenamefont {de~Juan}\ \emph {et~al.}(2014)\citenamefont
  {de~Juan}, \citenamefont {R\"uegg},\ and\ \citenamefont
  {Lee}}]{PhysRevB.89.161117}%
  \BibitemOpen
  \bibfield  {author} {\bibinfo {author} {\bibfnamefont {Fernando}\
  \bibnamefont {de~Juan}}, \bibinfo {author} {\bibfnamefont {Andreas}\
  \bibnamefont {R\"uegg}}, \ and\ \bibinfo {author} {\bibfnamefont {Dung-Hai}\
  \bibnamefont {Lee}},\ }\bibfield  {title} {\enquote {\bibinfo {title}
  {Bulk-defect correspondence in particle-hole symmetric insulators and
  semimetals},}\ }\href {\doibase 10.1103/PhysRevB.89.161117} {\bibfield
  {journal} {\bibinfo  {journal} {Phys. Rev. B}\ }\textbf {\bibinfo {volume}
  {89}},\ \bibinfo {pages} {161117} (\bibinfo {year} {2014})}\BibitemShut
  {NoStop}%
\bibitem [{\citenamefont {Rhim}\ \emph {et~al.}(2017)\citenamefont {Rhim},
  \citenamefont {Behrends},\ and\ \citenamefont
  {Bardarson}}]{PhysRevB.95.035421}%
  \BibitemOpen
  \bibfield  {author} {\bibinfo {author} {\bibfnamefont {Jun-Won}\ \bibnamefont
  {Rhim}}, \bibinfo {author} {\bibfnamefont {Jan}\ \bibnamefont {Behrends}}, \
  and\ \bibinfo {author} {\bibfnamefont {Jens~H.}\ \bibnamefont {Bardarson}},\
  }\bibfield  {title} {\enquote {\bibinfo {title} {Bulk-boundary correspondence
  from the intercellular zak phase},}\ }\href {\doibase
  10.1103/PhysRevB.95.035421} {\bibfield  {journal} {\bibinfo  {journal} {Phys.
  Rev. B}\ }\textbf {\bibinfo {volume} {95}},\ \bibinfo {pages} {035421}
  (\bibinfo {year} {2017})}\BibitemShut {NoStop}%
\bibitem [{\citenamefont {van Miert}\ and\ \citenamefont
  {Ortix}(2017)}]{PhysRevB.96.235130}%
  \BibitemOpen
  \bibfield  {author} {\bibinfo {author} {\bibfnamefont {Guido}\ \bibnamefont
  {van Miert}}\ and\ \bibinfo {author} {\bibfnamefont {Carmine}\ \bibnamefont
  {Ortix}},\ }\bibfield  {title} {\enquote {\bibinfo {title} {Excess charges as
  a probe of one-dimensional topological crystalline insulating phases},}\
  }\href {\doibase 10.1103/PhysRevB.96.235130} {\bibfield  {journal} {\bibinfo
  {journal} {Phys. Rev. B}\ }\textbf {\bibinfo {volume} {96}},\ \bibinfo
  {pages} {235130} (\bibinfo {year} {2017})}\BibitemShut {NoStop}%
\bibitem [{\citenamefont {Watanabe}\ and\ \citenamefont
  {Oshikawa}(2018)}]{PhysRevX.8.021065}%
  \BibitemOpen
  \bibfield  {author} {\bibinfo {author} {\bibfnamefont {Haruki}\ \bibnamefont
  {Watanabe}}\ and\ \bibinfo {author} {\bibfnamefont {Masaki}\ \bibnamefont
  {Oshikawa}},\ }\bibfield  {title} {\enquote {\bibinfo {title} {Inequivalent
  berry phases for the bulk polarization},}\ }\href {\doibase
  10.1103/PhysRevX.8.021065} {\bibfield  {journal} {\bibinfo  {journal} {Phys.
  Rev. X}\ }\textbf {\bibinfo {volume} {8}},\ \bibinfo {pages} {021065}
  (\bibinfo {year} {2018})}\BibitemShut {NoStop}%
\bibitem [{\citenamefont {Po}\ \emph {et~al.}(2017)\citenamefont {Po},
  \citenamefont {Vishwanath},\ and\ \citenamefont {Watanabe}}]{po2017symmetry}%
  \BibitemOpen
  \bibfield  {author} {\bibinfo {author} {\bibfnamefont {H.~C.}\ \bibnamefont
  {Po}}, \bibinfo {author} {\bibfnamefont {A.}~\bibnamefont {Vishwanath}}, \
  and\ \bibinfo {author} {\bibfnamefont {H.}~\bibnamefont {Watanabe}},\
  }\bibfield  {title} {\enquote {\bibinfo {title} {Symmetry-based indicators of
  band topology in the 230 space groups},}\ }\href
  {https://doi.org/10.1038/s41467-017-00133-2} {\bibfield  {journal} {\bibinfo
  {journal} {Nat. Commun.}\ }\textbf {\bibinfo {volume} {8}},\ \bibinfo {pages}
  {50} (\bibinfo {year} {2017})}\BibitemShut {NoStop}%
\bibitem [{\citenamefont {Bradlyn}\ \emph {et~al.}(2017)\citenamefont
  {Bradlyn}, \citenamefont {Elcoro}, \citenamefont {Cano}, \citenamefont
  {Vergniory}, \citenamefont {Wang}, \citenamefont {Felser}, \citenamefont
  {Aroyo},\ and\ \citenamefont {Bernevig}}]{bradlyn2017topological_nature}%
  \BibitemOpen
  \bibfield  {author} {\bibinfo {author} {\bibfnamefont {B.}~\bibnamefont
  {Bradlyn}}, \bibinfo {author} {\bibfnamefont {L.}~\bibnamefont {Elcoro}},
  \bibinfo {author} {\bibfnamefont {J.}~\bibnamefont {Cano}}, \bibinfo {author}
  {\bibfnamefont {M.~G.}\ \bibnamefont {Vergniory}}, \bibinfo {author}
  {\bibfnamefont {Z.}~\bibnamefont {Wang}}, \bibinfo {author} {\bibfnamefont
  {C.}~\bibnamefont {Felser}}, \bibinfo {author} {\bibfnamefont {M.~I.}\
  \bibnamefont {Aroyo}}, \ and\ \bibinfo {author} {\bibfnamefont {B.~A.}\
  \bibnamefont {Bernevig}},\ }\bibfield  {title} {\enquote {\bibinfo {title}
  {Topological quantum chemistry},}\ }\href
  {http://dx.doi.org/10.1038/nature23268} {\bibfield  {journal} {\bibinfo
  {journal} {Nature}\ }\textbf {\bibinfo {volume} {547}},\ \bibinfo {pages}
  {298--305} (\bibinfo {year} {2017})}\BibitemShut {NoStop}%
\bibitem [{\citenamefont {Ono}\ and\ \citenamefont
  {Watanabe}(2018)}]{PhysRevB.98.115150}%
  \BibitemOpen
  \bibfield  {author} {\bibinfo {author} {\bibfnamefont {Seishiro}\
  \bibnamefont {Ono}}\ and\ \bibinfo {author} {\bibfnamefont {Haruki}\
  \bibnamefont {Watanabe}},\ }\bibfield  {title} {\enquote {\bibinfo {title}
  {Unified understanding of symmetry indicators for all internal symmetry
  classes},}\ }\href {\doibase 10.1103/PhysRevB.98.115150} {\bibfield
  {journal} {\bibinfo  {journal} {Phys. Rev. B}\ }\textbf {\bibinfo {volume}
  {98}},\ \bibinfo {pages} {115150} (\bibinfo {year} {2018})}\BibitemShut
  {NoStop}%
\bibitem [{\citenamefont {Qi}\ \emph {et~al.}(2008)\citenamefont {Qi},
  \citenamefont {Hughes},\ and\ \citenamefont {Zhang}}]{PhysRevB.78.195424}%
  \BibitemOpen
  \bibfield  {author} {\bibinfo {author} {\bibfnamefont {X.-L.}\ \bibnamefont
  {Qi}}, \bibinfo {author} {\bibfnamefont {T.~L.}\ \bibnamefont {Hughes}}, \
  and\ \bibinfo {author} {\bibfnamefont {S.-C.}\ \bibnamefont {Zhang}},\
  }\bibfield  {title} {\enquote {\bibinfo {title} {Topological field theory of
  time-reversal invariant insulators},}\ }\href
  {https://link.aps.org/doi/10.1103/PhysRevB.78.195424} {\bibfield  {journal}
  {\bibinfo  {journal} {Phys. Rev. B}\ }\textbf {\bibinfo {volume} {78}},\
  \bibinfo {pages} {195424} (\bibinfo {year} {2008})}\BibitemShut {NoStop}%
\bibitem [{\citenamefont {Fu}\ and\ \citenamefont
  {Kane}(2007)}]{PhysRevB.76.045302}%
  \BibitemOpen
  \bibfield  {author} {\bibinfo {author} {\bibfnamefont {L.}~\bibnamefont
  {Fu}}\ and\ \bibinfo {author} {\bibfnamefont {C.~L.}\ \bibnamefont {Kane}},\
  }\bibfield  {title} {\enquote {\bibinfo {title} {Topological insulators with
  inversion symmetry},}\ }\href
  {http://link.aps.org/doi/10.1103/PhysRevB.76.045302} {\bibfield  {journal}
  {\bibinfo  {journal} {Phys. Rev. B}\ }\textbf {\bibinfo {volume} {76}},\
  \bibinfo {pages} {045302} (\bibinfo {year} {2007})}\BibitemShut {NoStop}%
\end{thebibliography}
%

\end{document}